\documentstyle[11pt,preprint,prd,aps]{revtex}

\tighten
\begin{document}

\draft

\title{Implications of Yukawa unification
for the Higgs sector\\
in supersymmetric grand-unified models}
\author{Paul Langacker and Nir Polonsky}
\address{Department of Physics, University of Pennsylvania,
Philadelphia, Pennsylvania, 19104, USA}

\date{February 1994, UPR-0594T}

\maketitle

\begin{abstract}
The $SU(5)$ unification-scale relation $h_{b} = h_{\tau}$
between the $b$-quark and $\tau$-lepton Yukawa couplings
severely constrains
$\tan\beta$ and the $t$-quark mass
(even more so if $h_{t}=h_{b} = h_{\tau}$ holds)
in supersymmetric models.
We examine the implications of these
constraints for the Higgs sector assuming
universal soft breaking terms, and emphasize
that both of these relations  impose
unique characteristics in terms of symmetries and of the spectrum.
We further study
the $\tan\beta \approx 1$
scenario, which is suggested by $h_{b} = h_{\tau}$,
and, in particular,  the loop-induced
mass of the light Higgs boson.
We compare the effective potential and renormalization
group methods and stress the  two-loop
ambiguities in the calculation of the mass.
These and a large enhancement
to the loop correction
due to $t$-scalar left-right mixing
considerably weaken the upper bound on the
mass of the light Higgs boson that has been reported.
Nevertheless,
we find that for this scenario the Higgs boson
is probably lighter
than 110 GeV, and typically lighter than 100 GeV.
Thus, it is in the mass range that
may be relevant for LEPII.
Our numerical results are presented in a self-contained
manner in section \ref{sec:s5}.
In separate appendices we discuss the global
symmetries of the Higgs potential,
the issue of false (color-breaking) vacua,
which may be important for $\tan\beta \approx 1$,
two-loop calculations,
and the effect of an additional Higgs singlet.
We show that
the approximate constraints that are often used
to eliminate color-breaking vacua
are not always relevant.
\end{abstract}

\section{introduction}
\label{sec:s1}

Supersymmetric grand-unified theories \cite{sgut,review} (SGUT),
though far from being proved,
are an attractive framework on both observational
and theoretical grounds. (Their relation to string theory
remains at present poorly understood.)
In the  minimal models, which we assume hereafter,  there is
a grand desert between the weak and the unification
scales ($M_{Z} \lesssim Q \lesssim $ TeV
and $M_{G} \approx 10^{16} - 10^{17}$ GeV,
respectively).
At the unification point $M_{G}$
\begin{equation}
\alpha_{1}(M_{G}) = \alpha_{2}(M_{G}) = \alpha_{3}(M_{G}) \equiv \alpha_{G}
\label{cc}
\end{equation}
(where $\alpha_{1} = \frac{5}{3}\alpha_{Y}$ and $\alpha_{G} \approx 0.04$),
and, depending on the
grand unifying group representations of the Higgs superfields that couple
to matter, certain relations between the standard model Yukawa couplings
hold. In the minimal models
based on $SU(5)$, $SO(10)$, and $E_{6}$
these representations are the
fundamental ones, and one obtains \cite{hbhtau}
\begin{equation}
h_{\tau}(M_{G})= h_{b}(M_{G}).
\label{yuk}
\end{equation}
For some models based on $SO(10)$ (or larger groups)
one has the stronger prediction
\begin{equation}
h_{\tau}(M_{G})= h_{b}(M_{G})= h_{t}(M_{G}),
\label{yuk2}
\end{equation}
but this only holds if one makes the additional assumption that the
masses are generated by a single complex Higgs {\bf10}-plet.
($h_{\tau,\,b,\,t}$ are the SM Yukawa couplings of the
$\tau,\,b,\,t$-superfields, respectively.)
One usually assumes that some perturbation modifies the
coupling or the masses of the two light families where, in principle,
similar relations should, but do not, hold\footnote{Since the Yukawa
couplings corresponding to the third family are much
larger than
those of the two light families, relation (\ref{yuk})
[or (\ref{yuk2})] will be only
slightly affected.}. (See also Appendix \ref{sec:app3}.)
Also, such relations often do not hold
in superstring unified models.

At the weak-scale the particle content is that of
the minimal supersymmetric standard model (MSSM), i.e.,
that of a two Higgs doublet model but with each bosonic (fermionic)
degree of freedom complemented by a fermionic (bosonic) one.
In particular, the Higgs sector contains three Goldstone bosons and five
physical degrees of freedom -- two CP even ($h^{0}$ and $H^{0}$) and one
CP odd ($A^{0}$) neutral, and one complex charged ($H^{+}$) Higgs bosons.
The symmetries and spectrum of the Higgs
scalars are strongly affected by the above GUT assumptions,
and are the subject of our present study. However, a review of
previous results in SGUT's, which
establish the base for our present work, is appropriate.

The consistency of (\ref{cc}) with the data was pointed out recently
in Ref. \cite{cc1} and that of (\ref{yuk}) and (\ref{yuk2})
in Ref. \cite{yuk1}.
Relation (\ref{cc}), however, holds only if there is exactly
one pair of Higgs doublets
(but any number of SM singlets is allowed).
We previously investigated the status of coupling constant unification
in great detail \cite{us1} where we considered non-trivial
matching conditions\footnote{Heavy degrees of freedom (which are model
dependent) are integrated out of the effective theory at $M_{G}$ and
some modification to the naive boundary conditions in (\ref{cc}) and
(\ref{yuk}) [or (\ref{yuk2})] is required, in general.
The appropriate matching functions may also include
corrections from nonrenormalizable operators since
$M_{G}$ is not too far from the Planck scale.
Similarly, integrating out the MSSM new particle spectrum
will correct the boundary conditions at $M_{Z}$.}
that perturb relation (\ref{cc}) in a model dependent
way. We concluded that no significant constraints on the MSSM parameters
can be deduced just by that relation. However, we find\footnote{
Eq. (\ref{als}) updates the results in \cite{us1} for the effect
of more recent precision electroweak data.}
that (\ref{cc}) predicts for the strong coupling
[using $\alpha(M_{Z})^{-1} = 127.9 \pm 0.1$ \cite{sirlin}
and $s^{2}(M_{Z}) = 0.2324 \pm 0.0003$
(for $m_{t}^{pole} = 143$ GeV)
for the
($\overline{{\mbox{MS}}}$)
fine-structure constant and weak angle, respectively]
\begin{equation}
\alpha_{s}(M_{Z}) = 0.125 \pm 0.001 \pm 0.008 + H_{\alpha_{s}}
+ 3.2 \times 10^{-7}{\mbox{GeV}}^{-2}[(m_{t}^{pole})^{2}
- (143\,{\mbox{GeV}})^{2}]\;,
\label{als}
\end{equation}
where the $\pm 0.008$ ($\pm 0.001$)  uncertainty in (\ref{als})
is theoretical (due to the input parameter error bars).
The former is due to the unknown values of the matching conditions
(i.e., threshold and nonrenormalizable-operator effects)
mentioned above.
The function $H_{\alpha_{s}}$ is a correction from Yukawa interactions.
It is negative but negligible, unless some
Yukawa couplings are $ \gtrsim 1$,
for which  we obtain $H_{\alpha_{s}} \approx - (0.001 - 0.003)$.
The quadratic $m_{t}^{pole}$ dependence is induced
by the correlation between
the weak angle extracted from the data
and the $t$-quark pole mass, $m_{t}^{pole}$.
Thus, typically one predicts $\alpha_{s}(M_{Z}) \gtrsim 0.12$
with a significant dependence on $m_{t}^{pole}$.
This observation is of importance when simultaneously considering
(\ref{cc}) and (\ref{yuk}) [or (\ref{yuk2})] to predict the
$b$-quark ($\overline{{\mbox{MS}}}$ running) mass $m_{b}$
[and/or the $t$-quark mass, if using (\ref{yuk2})]
in terms of the $\tau$-lepton mass.
We find [using, e.g., Eq. (\ref{als})
but with a more careful treatment of theoretical uncertainties
to the $b$-quark mass prediction]
that only a small region in the $\tan\beta - m_{t}^{pole}$ plane
is allowed by the experimental range of $m_{b}$ \cite{us2},
where $\tan\beta = {\nu_{up}/\nu_{down}}$ is the ratio of the two
Higgs doublet expectation values.
This holds even when non-trivial matching conditions
at both scales are considered.

\begin{figure}
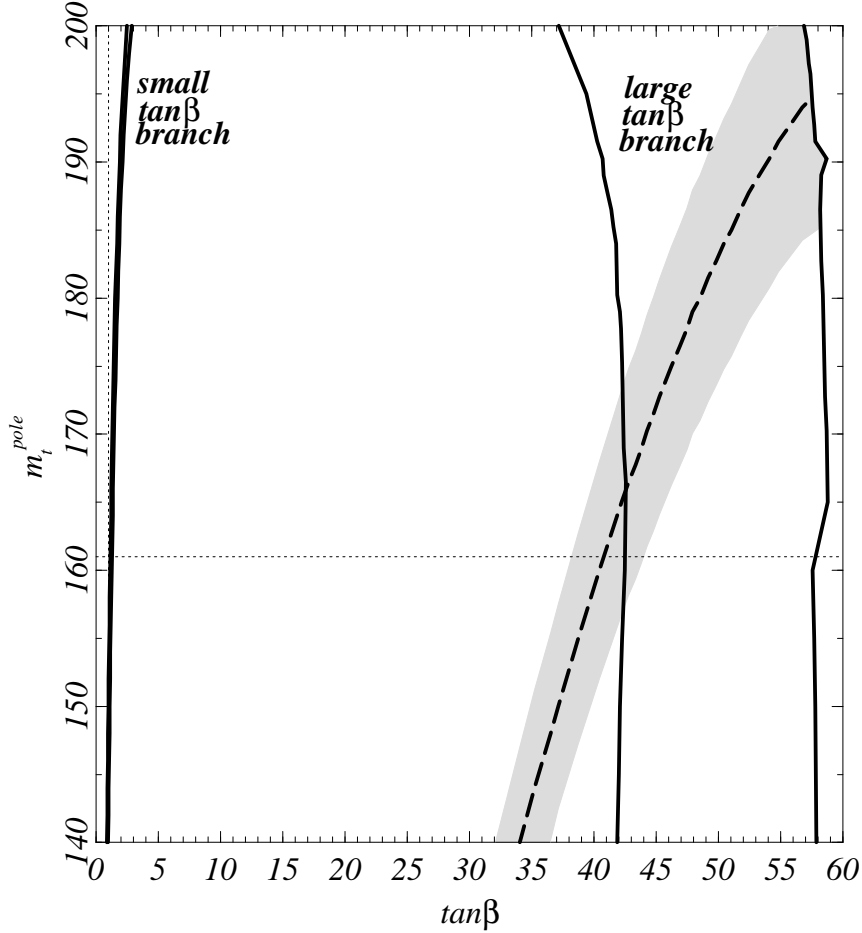

\caption{
The $t$-quark pole mass (in GeV)
 - $\tan\beta$ plane is divided
into five different regions.
Two areas (low- and high-$\tan\beta$ branches)
are consistent with perturbative
two-Yukawa ($b$ and $\tau$) unification
and with an upper bound of $\frac{4.45}{0.85}$ GeV for the $b$-quark current
mass.
The 0.85 factor takes into account a $\sim \pm 15\%$
theoretical uncertainty in the
numerical calculation due to the matching conditions.
Between the two branches the $b$-quark
mass is too high. For a too low (high) $\tan\beta$,
the $t$-quark ($b$-quark) Yukawa coupling
diverges.
The region where all three (third-family) Yukawa couplings
unify (dashed line) intersects
the allowed high-$\tan\beta$ branch.
[The three-Yukawa region is calculated without imposing
any constraints on $m_{b}$, and is assigned
a $\sim 5\%$ uncertainty (shaded area)
from corrections to the $h_{t}/h_{b}$ ratio.]
The $t$-quark mass range suggested by the electroweak data
and the $\tan\beta = 1$ line
are indicated (dotted lines) for comparison.
}
\label{fig:f1}
\end{figure}

It is customary to assume that the MSSM spectrum is given to a good
approximation near $M_{G}$ by a small number of universal
supersymmetry-breaking soft parameters; i.e., a common scalar mass $m_{0}$,
trilinear and bilinear (dimension-one) couplings $A_{0}$ and $B_{0}$,
and a common gaugino mass
$M_{\frac{1}{2}}$; in addition to a
supersymmetric Higgs mass (i.e., the Higgsino mass),
$\mu_{0}$  \cite{review}.
We will adopt these assumptions below.
(For an alternative scenario, see, for example, Ref. \cite{horiz}.)
However, our conclusions so far are largely
(i.e., except for the the details of the matching conditions)
independent of those assumptions and
hold in any SGUT  scenario
with an approximate grand-desert in which the MSSM $\beta$-functions
are valid.
The universality  assumption just
strengthens the constraints on the
$\tan\beta - m_{t}^{pole}$ plane; i.e.,
assuming universal initial conditions and
requiring that the weak-scale MSSM (one-loop improved) scalar potential
is consistent with a broken $SU(2)\times U(1)$
symmetry, we have to further constrain
$\tan\beta \geq 1$ and $h_{t} \geq h_{b}$. We then find two allowed regions,
\begin{enumerate}
\item $\tan\beta = 1 + \Delta_{\beta}$, $140 \lesssim m_{t}^{pole}
\,(\lesssim 200)$ GeV, $h_{t} \approx 1 \gg h_{b};$
\item $\tan\beta \approx 50 \pm 10$,
$170 \pm 10 \lesssim m_{t}^{pole} \, (\lesssim 200)$ GeV,
$h_{t} \gtrsim  h_{b} \approx 1$.
\end{enumerate}
The first of these solutions is consistent only with $h_{b} = h_{\tau}$,
and has $\Delta_{\beta} \lesssim 1$ for $m_{t}^{pole} \lesssim 190$ GeV.
A part of the region which is described by the second solution
is consistent with the stronger prediction
$h_{\tau} = h_{b} = h_{t}$.
A third solution
with $3 \lesssim \tan\beta \lesssim 40$ exists for
$m_{t}^{pole} \approx 215 \pm 10$ GeV, which is the upper bound on
$m_{t}^{pole}$ if the model is to stay perturbative up to $M_{G}$.
However, such large values for $m_{t}$ are inconsistent with
precision electroweak data.
The allowed regions are illustrated in Fig.\ \ref{fig:f1}.

These results are in good agreement with those of Ref.
\cite{mad1,maxp,mad2,hall,mich}, with any small differences due
to different treatments of matching conditions and of the
coupling constant unification condition.
Arbitrarily relaxing both condition (\ref{cc}) and (\ref{yuk})
independently
[e.g., $\alpha_{s}(M_{Z}) = 0.11$ and $h_{b}(M_{G}) = 0.85h_{\tau}(M_{G})$]
could invalidate our conclusions.
However, the corrections to (\ref{cc}) and (\ref{yuk})
are strongly correlated, and when
treated as such the strong constraints on the
$\tan\beta - m_{t}^{pole}$ plane hold\footnote{That is unless one assumes
a conspiracy of various corrections that are otherwise independent.}.
In particular, the
$m_{b}/m_{\tau}$ ratio is relatively insensitive
to the details of the decoupling of the low-scale spectrum \cite{us2},
and our conclusions would  not
be significantly modified
if those thresholds were treated differently
than in our (semi-analytic) approximation
(where we used six effective mass parameters).
Let us stress that any treatment of theoretical
uncertainties would be lacking unless the potentially
important corrections at the high scale are accounted for
in one form or another.
These issues are summarized in Ref. \cite{us3,us4}, where the second
reference updates some of the results of Ref. \cite{us1,us2}.
Here, relevant results are again updated using a preliminary
analysis of most recent precision data.

A two-parameter fit [allowing the light (SM-like)
Higgs boson mass,
$m_{h^{0}}$, to vary from
$50 - 150$ GeV with a central value $m_{h^{0}} = M_{Z}$] to all $Z$, $W$,
and neutral-current data yields
\begin{equation}
s^{2}(M_{Z}) = 0.2324 \pm  0.0003 - 0.92\times 10^{-7}{\mbox{ GeV}}^{-2}
[({m_{t}^{pole}})^{2} - {(143 {\mbox{ GeV}})}^{2}],
\label{weakangle}
\end{equation}
\begin{equation}
m_{t}^{pole} = 143^{+17}_{-19} + 12.5\ln{\frac{m_{h^{0}}}{M_{Z}}}
{\mbox{ GeV,}}
\label{mtop}
\end{equation}
which in turn implies $\tan\beta \lesssim 1.3$ (at one s.d.).
Eqs. (\ref{weakangle}) and (\ref{mtop}) update Ref. \cite{tasi}
where $m_{t}^{pole} = 134^{+23}_{-28} + 12.5\ln{\frac{m_{h^{0}}}{M_{Z}}}$
GeV was given (which also implied $\tan\beta \lesssim 1.3$).

It was pointed out by Barger et al. \cite{mad2}
that if indeed $m_{t}^{pole} \lesssim 160$ GeV (as suggested by
the range given in Ref.\cite{tasi} and here)
then $m_{h^{0}}$ [in case $(1)$]
is determined by the magnitude of the loop
corrections, $\Delta_{h^{0}}$, and is $\lesssim 85$ GeV.
If so, the $b$-quark mass prediction in minimal $SU(5)$
(or similar) SGUT constrains
the SM Higgs boson mass more significantly than the general
MSSM (triviality) upper bound  of $\sim 130$ GeV \cite{sher,madrid,mich1}.
We agree qualitatively with this conclusion, but argue
that two-loop ambiguities and important left-right $t$-scalar
mixing effects,
in addition to the higher $m_{t}^{pole}$ range,
weaken the bound to $m_{h^{0}} \lesssim 110$ GeV.

Below,
we will establish the basis for and then pursue
further numerical studies of the prediction for
$m_{h^{0}}$ and its upper bound in
the $\tan\beta \approx 1$ scenario.
Rather than approximating the one-loop correction
we will calculate $m_{h^{0}}$ for a given point
in the parameter space
using the supersymmetric spectrum (calculated numerically),
and use monte-carlo routines to study the upper bound.
In section \ref{sec:s2} we briefly review the MSSM Higgs potential
and its minimization conditions. We  specialize the
discussion to the minimal SGUT scenarios, cases $(1)$ and $(2)$,
and emphasize that the Higgs potential exhibits approximate
global symmetries in these cases, on which we elaborate
in Appendix \ref{sec:app1}. We concentrate thereafter on the
$\tan\beta \approx 1$ scenario. Some general features
are described in section \ref{sec:s3}. In particular, the large Higgsino
mass parameter can generate false vacuum solutions
with broken color and charge. We discuss that issue
in greater detail in Appendix \ref{sec:app2}.
The calculation of the light Higgs boson mass in both the
effective potential and renormalization group methods is discussed in
section \ref{sec:s4}, where we use the comparison between
the two methods to study the two-loop ambiguities in the calculation.
The details of the two methods, as well as two-loop
calculations, are further discussed in Appendix \ref{sec:app4}.
Section \ref{sec:s5} is reserved for further numerical studies of
the Higgs boson mass, and we show that
because of two-loop ambiguities and $\tilde{t}_{L}-\tilde{t}_{R}$
mixing $m_{h^{0}}$ can easily
exceed 85 GeV but, unless $m_{t}^{pole}$ is much heavier than the
range suggested by (\ref{mtop}), it remains in the range
that may be relevant for LEPII.
There is even some possibility that it is in
the range still relevant for LEPI.
We also discuss the relation between
$t$-scalar mixing and color breaking.
In particular, parameter ranges which correspond
to large mixing (and large $m_{h^{0}}$) often have
an unacceptable color-breaking global minimum,
but not always.
The discussion in section \ref{sec:s5}
is to a great extent independent of the other sections
and can be read on its own. We will summarize our conclusions in section
\ref{sec:s6}, where we also examine the implications
for $m_{t}^{pole}$. In Appendix \ref{sec:app3}
we point out simple extensions of either the MSSM
or the GUT in which our (constrained) analysis does not apply.

Below, we point out that
(negative-energy) color-breaking local minima of the
full scalar potential are generic.
Approximate analytic
constraints that are often used in the literature
attempt to eliminate not only global but also local
color-breaking minima, and are thus too strong.
Furthermore, they are designed
to eliminate only certain types of color-breaking minima,
and are thus too weak. This is explicitly
demonstrated in Appendix \ref{sec:app2}, where
we also compare analytic and numerical treatments of the problem.

We follow Ref. \cite{us2} in calculating the
couplings (and the unification point), and Ref. \cite{ep2}
in treating the one-loop effective potential correction
$\Delta V$, including contributions from all sectors.
The boundary conditions at $M_{G}$
for $0 \leq m_{0} \leq 500$ GeV,  $|A_{0}| \leq 3m_{0} $,
and $50 \leq M_{\frac{1}{2}} \leq 350$ GeV,
are picked at random.
$m_{t}^{pole}$
values are picked at random
but with a gaussian distribution defined by (\ref{mtop}).
For a given $m_{t}^{pole}$,
$\tan\beta(M_{Z})$ is picked at random within its allowed region
[case $(1)$].
Values of $m_{t}^{pole} < 155 $ GeV are discarded
so that $\tan\beta \gtrsim 1.1$ and the divergent limit
is not reached (see below).
We then solve iteratively for $\mu^{2}_{0}$ (and $B_{0}$).
[The relevant renormalization group equations
(RGE's) have been given by various authors \cite{rge}.]
The sign of $\mu_{0}$, which is a RG invariant, is picked at random as well.
We explicitly verify that the potential is bounded from below;
$SU(2) \times U(1)$ is properly broken;
the lightest supersymmetric particle (LSP) is neutral
(in practice, it is a neutralino);
and  that all the squared masses which
correspond to physical scalars are positive before a point
is accepted.
We then find, in general,  that we are in agreement with lower
bounds coming from direct searches.
(e.g., values of $M_{\frac{1}{2}} \lesssim 100$ GeV, which would imply
a too light chargino, are already excluded in our case by a tachionic
$t$-scalar.) We do not impose a lower bound on $m_{h^{0}}$, which
is the subject of our investigation.
Our treatment of color and/or charge breaking (CCB) minima is discussed
in Appendix \ref{sec:app2}.

\section{The weak-scale Higgs sector}
\label{sec:s2}
The Higgs part of the MSSM (weak-scale) scalar potential
reads \cite{higgs,sher}
\begin{eqnarray}
V(H_{1},\,H_{2}) = (m_{H_{1}}^{2} + \mu^{2})|H_{1}|^{2} +
(m_{H_{2}}^{2} + \mu^{2})|H_{2}|^{2}  & & \nonumber \\
+ B\mu(H_{1}H_{2} + h.c.)
+\frac{{\lambda}^{\mbox{\tiny MSSM}}}{2}(|H_{2}|^{2} - |H_{1}|^{2})^{2}
+\Delta V, & &
\label{pot}
\end{eqnarray}
where $m_{H_{1}}^{2}$, $m_{H_{2}}^{2}$, and $B$ ($\mu$)
are the soft (supersymmetric) mass parameters
renormalized down to the weak scale, $m_{3}^{2} \equiv B\mu < 0$,
$\lambda^{\mbox{\tiny MSSM}} = \frac{g_{2}^{2} + \frac{3}{5}g_{1}^{2}}{4}$,
and we suppress $SU(2)$ indices.
The one-loop correction \cite{ep,sher}
$\Delta V = \Delta V^{one-loop}$
(which, in fact, is a threshold correction
to the one-loop improved tree-level potential)
can be absorbed to a good approximation
in redefinitions of the tree-level parameters \cite{ep2,epexamp}.

A broken $SU(2)\times U(1)$
(along with the constraint
$m_{H_{1}}^{2} + m_{H_{2}}^{2} + 2\mu^{2} \geq 2|m_{3}^{2}|$
from vacuum stability)
requires \cite{higgs,sher}
\begin{equation}
(m_{H_{1}}^{2} + \mu^{2})(m_{H_{2}}^{2} + \mu^{2}) \leq |m_{3}^{2}|^{2},
\label{cond}
\end{equation}
and the minimization conditions then give \cite{higgs,sher}
\begin{mathletters}
\label{min}
\begin{equation}
\mu^{2} = \frac{m_{H_{1}}^{2} - m_{H_{2}}^{2}\tan^{2}\beta}{\tan^{2}\beta - 1}
-\frac{1}{2}M_{Z}^{2},
\label{min1}
\end{equation}
\begin{equation}
{m_{3}^{2}} = -\frac{1}{2}\sin2\beta\left[
{m_{H_{1}}^{2} + m_{H_{2}}^{2} + 2\mu^{2}}\right].
\label{min2}
\end{equation}
\end{mathletters}
For $\tan\beta \rightarrow 1$ one has $h_{t} \gg h_{b}$,
and hence cannot have $m_{H_{1}}^{2} = m_{H_{2}}^{2}$
(assuming universal initial conditions).
Thus, $|\mu| \rightarrow \infty$
in that limit [case $(1)$], and the $SU(2)\times U(1)$
breaking is driven by the $B\mu$ term.
In practice, we will
stay away from the divergent limit in case $(1)$ by taking
$\tan\beta \gtrsim 1.1$ \cite{ep2}.
We do not expect significantly different results for
$1\leq \tan\beta \leq 1.1$; i.e., the divergence is either
stabilized by model dependent finite-loop corrections
or would exclude that region (see also below).
For $\tan\beta \rightarrow \infty$ [case $(2)$]
one has $B\mu \rightarrow 0$
so that the symmetry breaking is driven by $m_{H_{2}}^{2} < 0$.

The $\tan\beta \rightarrow 1$ case corresponds to an approximate
$SU(2)_{L + R}$ custodial symmetry of the vacuum
\cite{cus,cus2}, which we further discuss
in Appendix \ref{sec:app1}.
The symmetry is broken at the loop level
so that one expects $\tan\beta$ slightly above
unity, in agreement with our cut $\tan\beta \gtrsim 1.1$.
As a result of the symmetry, the CP-even Higgs
mass matrix becomes
\begin{equation}
M^{2} \approx  \mu^{2} \times
\left( \begin{array}{cc}1 & -1 \\ -1 & 1\end{array} \right),
\nonumber
\end{equation}
and it has a massless tree-level eigenvalue, $m_{h^{0}}^{T} \approx 0$.
This is, of course, a well known result of the tree-level formula
\cite{higgs,sher}
\begin{equation}
{m_{h^{0}}^{T}}^{2} = \frac{1}{2}\left[
m_{A^{0}}^{2} + M_{Z}^{2} -
\sqrt{(m_{A^{0}}^{2} + M_{Z}^{2})^{2} -
4m_{A^{0}}^{2} M_{Z}^{2}\cos^{2}2\beta \,} \right]
\label{tree}
\end{equation}
when taking $\beta = \frac{\pi}{4}$.
The mass is then determined by the loop correction
$m_{h^{0}} \approx \Delta_{h^{0}} \propto h_{t}m_{t}$
and is further studied in sections \ref{sec:s4} and \ref{sec:s5}.
On the other hand, in case $(2)$ $\beta \rightarrow  \frac{\pi}{2}$
and we have $m_{h^{0}}^{T} \approx M_{Z}$ (assuming
$m_{A^{0}} \geq M_{Z}$).
When adding the loop
correction\footnote{$\Delta_{h^{0}}$ is enhanced in this case
by the large $m_{t}$ and the heavy $t$-scalars.}
$m_{h^{0}} \lesssim \sqrt{2}M_{Z} \approx 130$ GeV.
The heavier CP-even Higgs boson mass eigenvalue
equals approximately $\sqrt{2}|\mu|$
in case $(1)$ and $m_{H^{0}} \approx m_{A^{0}}$ in case $(2)$.
The loop corrections are less relevant here as typically
$m_{H^{0}}^{2} \gg \Delta_{H^{0}}^{2}$
(in particular, when  $|\mu| \rightarrow \infty$).

The custodial symmetry (or the large $\mu$ parameter)
dictates in case $(1)$ a degeneracy
$m_{A^{0}} \approx m_{H^{0}} \approx
m_{H^{+}} \approx  \sqrt{2}|\mu|$.
[The tree-level corrections to that relation
are of order $(M_{W,\,Z}/m_{A^{0}})^{2}$.]
That is, at a scale $Q \approx \sqrt{2}|\mu| \approx 2$ TeV
the heavy Higgs doublet $H$ is decoupled, and the
effective field theory below that scale
has only one
massless ($m_{h^{0}}^{T} \approx 0$) Coleman-Weinberg
SM-like ($\nu_{h^{0}} = \nu$) Higgs doublet, $h$.
Unlike in our case, in general
$M_{Z}/m_{A^{0}}$ is not guaranteed
to be small, and the decoupling scale $Q$ is much lower,
typically below the scalar-quark thresholds.

The Higgs sector in case $(2)$ exhibits an
approximate $O_{4} \times O_{4}$ symmetry
(see Appendix \ref{sec:app1}). This is
a special case of the $m_{3}^{2} \rightarrow 0$ case
discussed in Ref. \cite{cus3}.
To obtain  $m_{3}^{2} \approx 0$
(starting with universal initial conditions
and $h_{b} \approx h_{t}$)
one would require large $M_{\frac{1}{2}}$ and $|\mu_{0}|$,
which enable one to obtain the desired limit by
adjusting large cancellations \cite{warsaw,bartol,maxp3}.
Once expectation values are acquired, $A^{0}$
and $H^{+}$ are massive pseudo-Goldstone bosons
of the broken $O_{4}$ symmetry. However, they are typically heavy
(for $\nu_{down} = 0$ the $O_{4}$ symmetry is restored)
but lighter than the scalar quarks in this scenario
(whose mass $m_{\tilde{q}} \approx
\sqrt{m_{0}^{2} + 6M_{\frac{1}{2}}^{2}} \approx 1$ TeV).

Large $\tan\beta$ [case $(2)$] solutions are
discussed in Ref. \cite{warsaw,bartol,hall,nel,maxp3}.
They predict $m_{t}^{pole} \approx 180 \pm 15$ GeV
\cite{us2,hall}, heavy scalars, and contain potentially
troublesome loop corrections, e.g., to $m_{b}$,
$b \rightarrow s\gamma$ \cite{hall,nel}; and, in the
absence of light scalars to counterbalance large
$m_{t}$ contributions, also to $Z \rightarrow b\bar{b}$
(see Ref. \cite{bvertex}).
The former can eliminate the motivation for any Yukawa
coupling analysis.
A different
approach was recently presented in Ref. \cite{maxp3}
where those large finite-loop corrections
were used to construct models
consistent with universality, a broken $SU(2)\times U(1)$,
and $h_{t} = h_{b} = h_{\tau}$.
Finite superpartner ($\sim 1$ TeV) loops are adjusted to
diminish $m_{b}$ by the right amount ($\sim 1$ GeV), and
$m_{b}$ is strongly dependent on the values of
the soft parameters, a possible but undesirable
situation. Much lower values of $m_{t}^{pole}$ are
then preferred. Diminishing $m_{t}$ may rectify the situation
with $Z \rightarrow b\bar{b}$, but the prediction for the
$b \rightarrow s\gamma$
branching ratio can be significantly modified
by the same superpartner loops \cite{hall}.
In contrast, the $\tan\beta \approx 1$
[case $(1)$] solution
constrains only $m_{t}^{pole} \gtrsim 140$ GeV
($m_{t}^{pole} \gtrsim 155$ GeV when imposing
the $\tan\beta \gtrsim 1.1$ cut),
is in better agreement with (\ref{mtop}), and predicts
a light Higgs boson ($m_{h^{0}} \lesssim 110$ GeV).
Cases $(1)$ and $(2)$ are summarized and compared
in Table \ref{table:t1}.
We hereafter study the $\tan\beta \approx 1$
[case $(1)$] solution only.
We assume that finite-loop
corrections to $m_{b}$ are at the most a few percent
which is, in general, the case away from
the large $\tan\beta$ limit, and
thus do not alter our results.
[The parameter space is sensitive to those corrections
only in case $(2)$.]

\section{The $\tan\beta \rightarrow 1$ scenario}
\label{sec:s3}

Our primary motivation to study the $\tan\beta \approx 1$ solution to the
MSSM is its consistency with condition (\ref{yuk}). The large  values
of $\alpha_{s}(M_{Z})$ implied by (\ref{cc}) [e.g., Eq. (\ref{als})]
generate (too) large positive loop corrections to $h_{b}/h_{\tau}$,
which need to be counterbalanced by large negative corrections
$\propto h_{t}^{2},\,h_{b}^{2}$.
The latter are sufficient only if some of the the Yukawa couplings
are large, i.e., near their unitarity upper bounds, which is the case
in the allowed regions.
[Also, $H_{\alpha_{s}}$ in (\ref{als}) becomes non-negligible
in that limit.]
The large values of the Yukawa couplings
can be understood in terms of a quasi-fixed point in the flow of
the respective RGE's \cite{hill}.
In particular, in case $(1)$ $h_{t}$ is near its quasi-fixed point.
The two issues, i.e., the consistency of $\tan\beta \approx 1 $
with (\ref{yuk}) and the $h_{t}$-RGE flow structure,
are strongly related.

The smallness of $\Delta_{\beta}$
(or alternatively, the large $|\mu|$)
can be understood in terms of an approximate
$SU(2)_{L} \times SU(2)_{R}$ symmetry in the Higgs potential,
which, if it exists at some scale, will be only slightly broken at $M_{Z}$,
as is shown in Appendix \ref{sec:app1}.
The $\tan\beta \approx 1$
scenario was previously studied in Ref. \cite{italy}, where it was
referred to as ``highest classical degeneracy'':
in the $\tan\beta = 1$ limit
condition (\ref{cond}) becomes
an equality. [In practice, we find that
${(m_{H_{1}}^{2}+\mu^{2})( m_{H_{2}}^{2} + \mu^{2})}/
{m_{3}^{4}} \gtrsim 0.99$
for $\Delta_{\beta} < 1$ and after the proper redefinitions.]
It was subsequently studied in Ref. \cite{haberdiaz} in the context
of a global MSSM (i.e., no high-scale assumptions),
and its consistency with limits on
$m_{h^{0}}$ was shown.
The $\tan\beta \approx 1$ solution suggested
by the minimal SGUT relation $h_{b} = h_{\tau}$
revived the interest in that corner of
parameter space. More recently Ref. \cite{maxp2,mad3,bartol2},
motivated by the quasi-fixed point prediction to $m_{t}$,
extended the discussion
from the Higgs sector to the full parameter space.

\begin{figure}
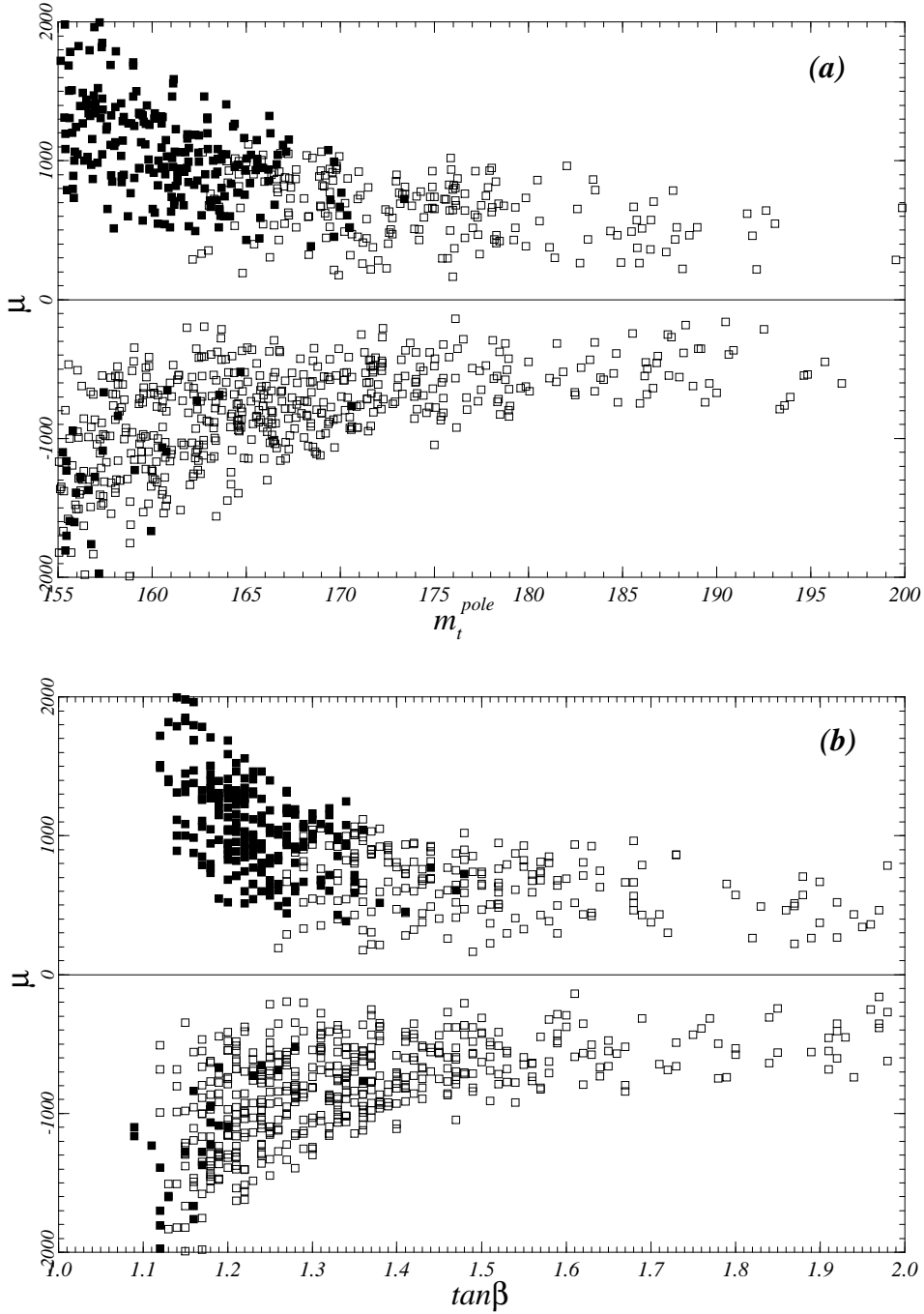

\caption{
The prediction for the Higgsino mass parameter $\mu$
(after absorbing $\Delta V$ in the redefined tree-level parameters)
as a function of ($a$) the $t$-quark pole mass and of
($b$) $\tan\beta$. (The $t$-quark mass  and $\tan\beta$
are strongly correlated.) Filled squares
indicate (probably unacceptable) points for which the
SM minimum is only a local minimum.
All masses are in GeV
}
\label{fig:new}
\end{figure}

In Figs.\ \ref{fig:new}a and b we show the prediction for
$\mu$ as a function of $m_{t}^{pole}$ and of $\tan\beta$,
respectively.
Typically $ |\mu| \sim 1$ TeV,
depending on $\Delta_{\beta}$ and on the soft parameter scale.
It is the large $|\mu|$ parameter that dictates the
characteristics of the
scenario (and not only for the Higgs sector).
Before further elaborating on the calculation
of $m_{h^{0}}$, let us briefly summarize some of the
features that appear in our numerical studies, and which are of
relevance for the discussion below.
Though we agree on the characteristics
with other authors,
our emphasis and interpretation are different.

\begin{enumerate}
\item
The Higgsino (with mass $\sim \mu$) is decoupled
from the (much lighter) gauginos. The bino and the wino are
the lightest neutralino and chargino, respectively.
(The former is the LSP which has a bino fraction near unity.)
The heavy Higgsino decouples from the $Z \rightarrow b\bar{b}$ vertex,
and thus, smaller values of $m_{t}^{pole}$ are
favored\footnote{The prediction for the
$Z \rightarrow b\bar{b}$ branching ratio decreases
with $m_{t}^{pole}$, and its measured value is $\sim 1.5$ s.d.
higher than the SM prediction with $m_{t}^{pole} \approx 140$ GeV.
The discrepancy grows with $m_{t}^{pole}$ \cite{jens}.}.
However, $m_{t}^{pole}$ and $\tan\beta$ are correlated,
and as $m_{t}^{pole}$ grows $|\mu|$ is diminished
(see Fig.\ \ref{fig:new}).
Thus, in principle,
Higgsino -- $t$-scalar loops can still counterbalance large
$m_{t}$ contributions
 to $Z \rightarrow b\bar{b}$ (see, for example,
Ref. \cite{bvertex}) if the $t$-scalar is light enough.
[A light $t$-scalar
is, however, less likely if  $|\mu|$ is small (see below) and
from the $\rho$-parameter.]
\item
The $t$-scalar mass-squared matrix is
(we take all parameters to be real)
\begin{equation}
\left( \begin{array}{cc}
m_{\tilde{t}_{L}}^{2}  &
m_{t}(A_{t} + \mu/\tan\beta) \\
m_{t}(A_{t} + \mu/\tan\beta) &
m_{\tilde{t}_{R}}^{2}
\end{array} \right),
\label{stop}
\end{equation}
and similarly for the superpartners of the other fermions.
The diagonal
elements correspond to the left and right-handed $t$-scalar
squared masses, $m_{\tilde{t}_{L,\,R}}^{2}$,
which consist of soft,
$F$, and $D$ terms.
$A_{t}$ is the trilinear soft parameter
preceeding the $h_{t}H_{2}\tilde{t}_{L}\tilde{t}_{R}$
term in the scalar potential, with $A_{t}(M_{G}) = A_{0}$.
The prediction for the diagonal terms has no
unique charachteristics in our case, and the left and right-handed
masses are in the $\sim 100$ GeV $- 1$ TeV range.
However, the mixing term is typically large
(unless the fermion mass is $\approx 0$
as is the case for the light families).
In the limit $|\mu| \rightarrow \infty$
there is a $\theta_{t} =  \frac{\pi}{4}$ left-right mixing
so that one of the
eigenstates would have a negative mass squared,
and the limit cannot be fully realized.
In practice, the soft terms and $m_{t}|\mu|$
are often of the same order of magnitude,
with a nearly degenerate mass-squared matrix.
Thus, the requirement
that (\ref{min}) defines a physical minimum
with no negative squared masses,
corresponding to physical scalars, constrains $\mu$.
Also, our previous comment regarding a light $t$-scalar and
a small $\mu$-parameter is clear by observation of (\ref{stop}).
It is interesting to note that
by fine adjustments of the soft parameters
[i.e., of the degeneracy in (\ref{stop})] one
may realize a scenario with a light ($\lesssim 45$ GeV)
$t$-scalar which is nearly
decoupled from the $Z$ (e.g., $\sin\theta_{t} \approx 0.8$).
(See, for example, Ref. \cite{lightstop}.)
However, such scenarios
are often associated with false vacua
(see below and Appendix \ref{sec:app2}).
\item
The absence of
a negative $t$-scalar squared mass determines a ($m_{0}$
and $\Delta_{\beta}$ dependent) lower bound on $M_{\frac{1}{2}}$.
In particular, ``no-scale'' (i.e., $m_{0} = A_{0} = 0$)
SGUT models are strongly constrained\footnote{
The strict no-scale assumption $B_{0} = m_{0} = A_{0} = 0$
is not consistent here. See Appendix \ref{sec:app1}.}
\footnote{No-scale models with $h_{b} = h_{\tau}$
were recently discussed in Ref. \cite{gunion}}.
Also, requiring that the light eigenstate of the
$\tau$-scalar mass-squared matrix (which carries charge)
is not the LSP
determines an upper bound on $M_{\frac{1}{2}}$
in the no-scale limit
(see also Kane et al. \cite{mich1}).
These point at a correlation between the bounds on the soft parameters.
A different class of correlations which take effect in the
$h_{t} \rightarrow h_{t}^{fixed}$ limit is discussed by Carena et al.
\cite{maxp2}.
Strong correlations (e.g., the $m_{t}^{pole} - \tan{\beta}$
correlation\footnote{
Note that even if we do not require $h_{b} = h_{\tau}$,
$m_{t}^{pole}$ and the lower bound on $\tan{\beta}$
are correlated due to unitarity considerations.}, the
$h_{t} \rightarrow h_{t}^{fixed}$ ones, and those
due to scalar mixings discussed here)
do not allow elimination
of the large-$|\mu|$ solutions (that correspond to
large left-right mixing enhancement to $m_{h^{0}}$ -- see below)
on the basis of fine-tuning arguments. The naive notion
of fine-tuning  becomes
obscure and misleading in the
presence of strong correlations.
(See also Ref. \cite{maxp2}.)
Rather, strong correlations
make the scenario quite predictive.
\item
We can rewrite the mixing term in (\ref{stop}) as
$h_{t}H_{2}^{0}\hat{\Gamma}$ where $H_{2}^{0} = \nu_{up}$.
$\hat{\Gamma}$ is strongly constrained by requiring that the
full scalar potential does not have a
color and/or charge breaking global (GCCB) minimum,
which is a different issue  than
whether the extremum defined by (\ref{min}) is indeed
a minimum.
The issue of false vacua is of particular interest
in the $\tan\beta \rightarrow 1$ scenario as
$V(H_{1},\,H_{2}) \rightarrow 0^{-}$ in that limit.
Below, we will consider that subject in some detail.
In particular, see  Appendix \ref{sec:app2}. We find
that the absence of negative energy CCB minima implies
the constraint, first described in Ref. \cite{ccb3},
\begin{equation}
\left( |A_{t}| + s|\mu| \right)^{2} \leq
2(m_{\tilde{t}_{L}}^{2} +m_{\tilde{t}_{R}}^{2}),
\label{ccb3a}
\end{equation}
where $s = A_{t}\mu/|A_{t}\mu|$,
$(|A_{t}| + s|\mu|)^{2}  = \hat{\Gamma}^{2}$  (for $\tan\beta = 1$),
and $m_{\tilde{t}_{L,\,R}}^{2}$ consist of only
the soft terms.
(Note the importance of the trilinear parameter
and of the relative sign.)
Our numerical studies imply that (\ref{ccb3a})
is (to a good approximation) a sufficient condition
(in that region of parameter space)
but is not necessary (i.e., it is too restrictive).
For example, more than $80\%$ of the points that are
inconsistent with that constraint for $m_{t}^{pole} \gtrsim 165$ GeV
correspond to CCB minima which are only local and are therefore safe.
In Fig.\ \ref{fig:new} only filled squares correspond to
(probably unacceptable) GCBB minima. Also
observe in Fig.\ \ref{fig:new} that
$\mu < 0$ is preferred by CCB constraints.
This is because typically $A_{t} \gg A_{0}$
for $A_{0} < 0$  so that
the contribution from a large
and positive $\mu$ cannot be canceled in (\ref{ccb3a})
(for our range of $A_{0}$).
\end{enumerate}

\section{The loop-induced mass}
\label{sec:s4}

Let us now proceed to discuss the mass of the SM-like
higgs boson. The upper bound on $m_{h^{0}}$ is given,
e.g., in Ref. \cite{tam1},
\begin{equation}
m_{h^{0}}^{2} \leq M_{Z}^{2}\cos^{2}2\beta
 + \frac{3\alpha m_{t}^{4}}{4\pi s^{2} (1 -s^{2})M_{Z}^{2}}
\left\{\ln\left(\frac{m_{\tilde{t}_{1}}^{2}m_{\tilde{t}_{2}}^{2}}
{m_{t}^{4}}\right) + \Delta_{\theta_{t}} \right\}
\label{mh}
\end{equation}
where
\begin{eqnarray}
&\Delta_{\theta_{t}} =
\left(m_{\tilde{t}_{1}}^{2}-m_{\tilde{t}_{2}}^{2}\right)
\frac{\sin^{2}2\theta_{t}}{2m_{t}^{2}}\ln\left(
\frac{m_{\tilde{t}_{1}}^{2}}{m_{\tilde{t}_{2}}^{2}}\right)
& \nonumber \\
&  + \left(m_{\tilde{t}_{1}}^{2}-m_{\tilde{t}_{2}}^{2}\right)^{2}
\left(\frac{\sin^{2}2\theta_{t}}{4m_{t}^{2}}\right)^{2}
\left[2 - \frac{m_{\tilde{t}_{1}}^{2}+m_{\tilde{t}_{2}}^{2}}
{m_{\tilde{t}_{1}}^{2}-m_{\tilde{t}_{2}}^{2}}
\ln\left(\frac{m_{\tilde{t}_{1}}^{2}}{m_{\tilde{t}_{2}}^{2}}\right)
\right], &
\label{mixing}
\end{eqnarray}
and where $m_{\tilde{t}_{i}}^{2}$ are the eigenvalues of the $t$-scalar
mass-squared matrix, $\theta_{t}$ is the mixing angle,
and we have neglected other loop contributions. (They  are included
in our numerical calculations below.)
The tree-level mass squared, ${m_{h^{0}}^{T}}^{2}$, and the loop correction,
$\Delta_{h^{0}}^{2}$, are bounded by the first
and second terms on the r.h.s. of Eq. (\ref{mh}), respectively.
In the absence of mixings $\Delta_{\theta_{t}} = 0$.
For $\tan\beta \rightarrow 1$ one obtains
$m_{h^{0}}^{T} \rightarrow 0$, and thus
$m_{h^{0}} \approx \Delta_{h^{0}}$.

The upper bound on $\Delta_{h^{0}}$ scales as $m_{t}^{4}$
and depends on the overall scale of the $t$-scalar mass,
the amount of left-right mixing, and the separation
$|m_{\tilde{t}_{1}}^{2}-m_{\tilde{t}_{2}}^{2}|$.
The large mixing and separation for $|\mu| \rightarrow \infty$
enhance  $\Delta_{h^{0}}$ significantly (typically
by $10 -20$ GeV) compared to the first term alone (which was all that was
included in Ref. \cite{mad2}).
Condition (\ref{yuk}) (and $h_{t}$ unitarity) correlate
$m_{t}^{pole}$ and $\tan\beta$ in case $(1)$ (see Ref. \cite{us2,us4}),
and $\Delta_{\beta}$ increases from
$\sim 0$ for $m_{t}^{pole} \sim 140$ GeV to
$\sim 1$ for $m_{t}^{pole} \sim 190$ GeV.
By increasing $\tan\beta$ one  gradually departs from
the $|\mu| \rightarrow \infty$ limit, and the
enhancement to $\Delta_{h^{0}}$ from left-right mixing
decreases. Thus, there is an interplay
between $\Delta_{\theta_{t}}$ and the overall factor of $m_{t}^{4}$.
(Of course, $m_{h^{0}}^{T}$ grows with $\tan\beta$
and hence with $m_{t}^{pole}$.)

As was pointed out in the previous section
(see also Appendix \ref {sec:app2}),
$\mu$ and the amount of mixing $\hat{\Gamma}$ are
subject to color and/or charge breaking (CCB) constraints.
In this section points which are consistent with (\ref{ccb3a}),
i.e., which have no negative-energy CCB minima,
are marked by
a circle  -- {\Large $\circ$}.
Points which are inconsistent with (\ref{ccb3a})
and therefore have either a local or global
negative-energy CCB minimum
are marked by a ``diamond''-- {\Large $\diamond$}.
The filled diamonds
correspond to a {\bf global} color breaking minimum
(as determined using numerical procedures),
which is (most likely) unacceptable. The open diamonds
represent CCB minima which are only local (i.e., above
the standard model minimum) and are therefore safe.
Constraint
(\ref{ccb3a}) is always evaluated at the $t$-scalar scale,
independent of the scale chosen for minimization.
The same is true for the numerical search of
global CCB (GCCB) minima.

Calculating $m_{{h}^{0}}$ is reduced
(in our case) to a great extent to calculating
$\Delta_{h^{0}}$, i.e., the deviation from what would be the case
if the matter were supersymmetric at the weak scale [e.g.,
$m_{{\tilde{t}}_{1}} = m_{{\tilde{t}}_{2}} = m_{t}$
in (\ref{mh}) and (\ref{mixing})].
Two standard ways to perform the calculation
are the effective potential method (EPM)
(e.g., see Ref. \cite{ez,tam1,brig,haberdiaz}) and the
renormalization group method (RGM) (e.g., see Ref.
\cite{haberham,ham,haberdiaz}). The two methods
correspond to a
``run and diagonalize'' and ``diagonalize and run''
algorithm, respectively.
The algorithms and their implementations are discussed
and compared in greater detail in Appendix \ref{sec:app4}.
Here we would like to demonstrate that the EP and RG methods
do agree (within their domains of validity), i.e.,
the results are not sensitive to the way one treats
the threshold corrections.
In particular, both methods exhibit a large
($10 - 20 \%$) two-loop ambiguity
from the presence of some two-loop residual
terms in the calculations (next-to-leading and leading
two-loop logarithms, as well as finite terms).
We explicitly identify  and discuss such terms in Appendix
\ref{sec:app4}.
The residual logarithmic two-loop terms can be described
in terms of a scale ambiguity (i.e., the scale dependence
of the one-loop calculation is of two-loop order).
Indeed, comparing the two methods
and studying that ambiguity
allow us to estimate the
two-loop correction. However, the
uncertainty remains.

In the EPM the ambiguity is explicitly related
to a scale dependence, as one is required to
specify a subtraction scale $Q$
(i.e., halt running at $Q$ and diagonalize mass matrices).
Indeed, including $\Delta V^{one-loop}(Q)$ corrects
for  one-loop leading logarithms, and thus
guarantees a scale independence of the calculation,
but only to that order (and up to wave-function
renormalization).
It is suggestive to take $Q=M_{Z}$, i.e., the scale
at which the physical inputs are given.
Another attractive choice is a scale at which
$\Delta V^{two-loop}(Q)$ is minimized.
We will not attempt to find such a scale
rigorously, but estimate $Q \approx 600$ GeV.
(One indeed expects such a $Q$ to be in the vicinity
of some average scalar-quark scale, as for
$\Delta V^{one-loop}$ in Ref. \cite{grz}.)
We will use the comparison
between the two methods to justify our choice.
(Of course, there is no one choice that is suitable for all
sets of parameters.)

In the RGM, which sums the leading
logarithms to all orders
if one integrates numerically,
the ambiguity is more apparent in
terms of orders in perturbation theory
(i.e., the order at which the summation
is truncated in approximate iterative solutions).
However, there
is also an explicit scale dependence from the
choice of decoupling scales for the heavy fields
(i.e., decouple heavy
eigenstates, calculate RGE's, and run to the
next decoupling scale).
We will decouple the heavy Higgs
doublet $H$ at $m_{A^{0}} \approx 2.1$ TeV and other fields at
lower thresholds\footnote{We checked that the details
of the choices for those intermediate scales
introduce a $\sim1\%$ uncertainty.}.
When decoupling $H$ we include appropriate matching conditions
\cite{ham}.

The EPM is more straightforward
to implement (and is therefore more common in the literature)
while the RGM requires a careful
consideration of decoupling scales and matching conditions.

\begin{figure}
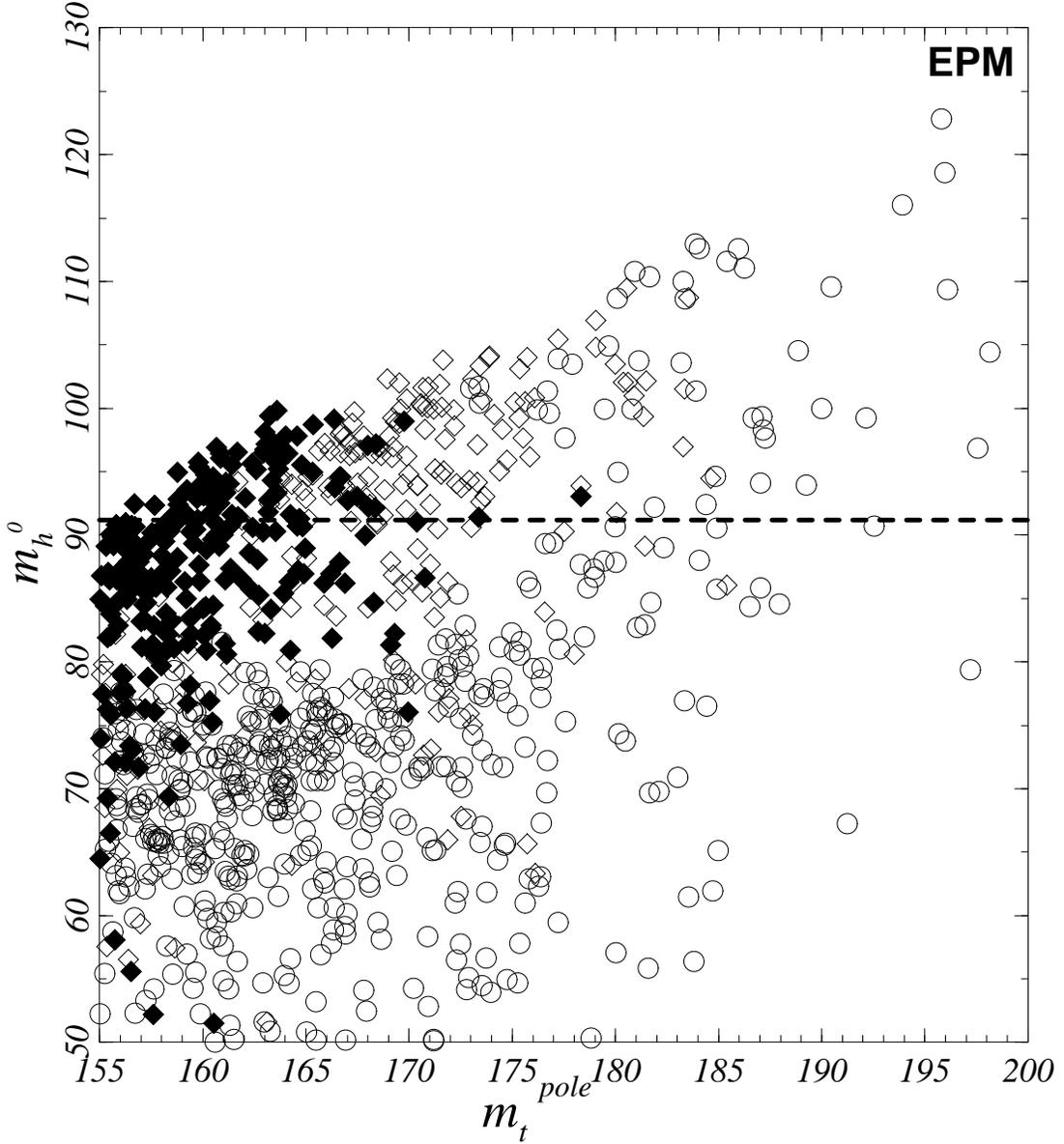

\caption{
A scatter plot of the Higgs boson mass (calculated in the
EPM with the subtraction scale $Q$ at the $Z$-pole) vs. the $t$-quark
pole mass. The $\circ$ indicate points which have no
(potentially dangerous) negative-energy color and/or charge breaking
minimum. The $\diamond$ have such a minimum but it is
a (safe) local one, i.e., it
lies above the standard-model minimum.
Points indicated by a filled diamond have an unacceptable
CCB global minimum.
The $Z$ mass is indicated for comparison
(dashed line). All masses are in GeV.
}
\label{fig:f2}
\end{figure}

\begin{figure}
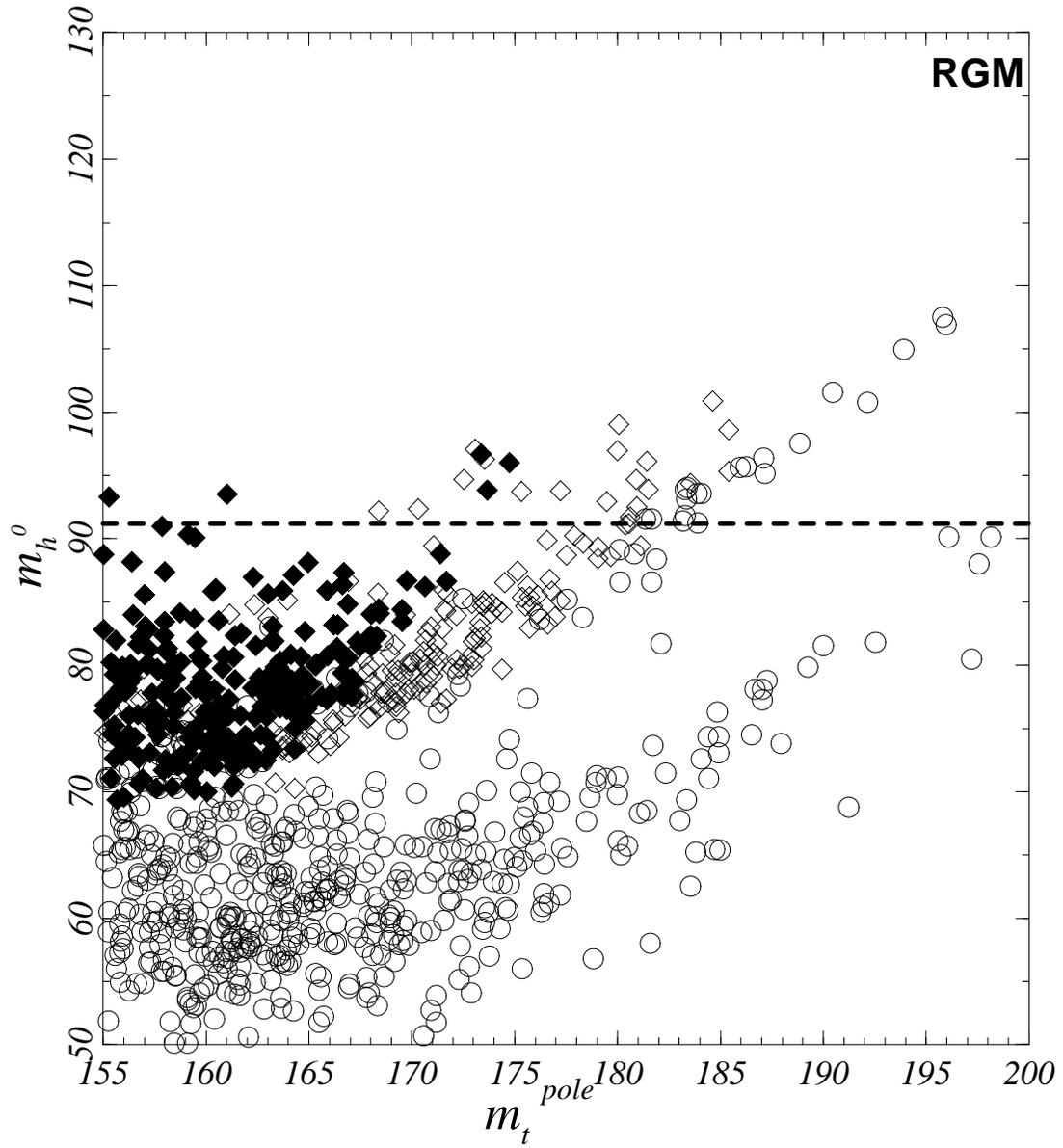

\caption{
Same as in Fig.\ 3 except the mass is calculated
using the RGM (numerically).
}
\label{fig:f3}
\end{figure}

The scatter plot Fig.\ \ref{fig:f2} displays the results of an EPM
calculation with $Q = M_{Z}$.
(When increasing $Q$ the soft mass parameters
and $m_{t}$ decrease so that the EPM $m_{h^{0}}$ is diminished.)
The scatter plot Fig.\ \ref{fig:f3} displays the results of a RGM calculation
(performed numerically).
Comparing them, one observes a $10-20$ GeV difference
between the respective
Higgs boson masses (for a given point in the
parameter space).
However, this is not due to the different choices
of an algorithm, but to the different
residual two-loop dependences in the two calculations.
To illustrate the maximum ambiguity
we used the $t$-quark pole (running) mass
in the calculations presented in Fig.\ \ref{fig:f2} (Fig.\ \ref{fig:f3}).
(The $\sim 5\%$ difference in the $t$-quark masses
is formally of two-loop order.)
In both plots one can observe a separation into a
heavier and a lighter Higgs boson branches. This is
due to the enhancement to $m_{h^{0}}$ in some cases
from ${\tilde{t}}_{L} -{\tilde{t}}_{R}$ mixing,
and is further discussed in the following section.

\begin{figure}
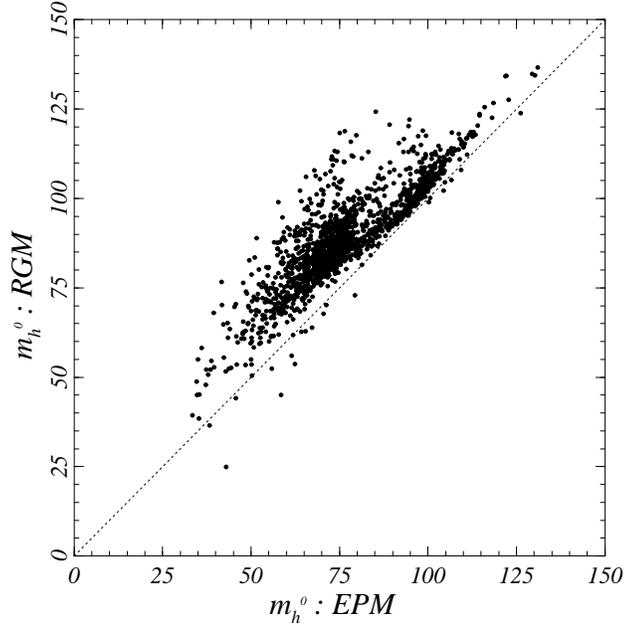

\caption{
The EPM (with the subtraction scale $Q$ at the $Z$-pole)
calculation of the Higgs boson mass
is compared with the mass calculated in the RGM
but using the Haber and  Hempfling  leading logarithm formula.
All masses are in GeV. Points corresponding to global GCCB
minima (indicated by a filled diamond in Figs.\ 3 and 4) are omitted.
}
\label{fig:f4}
\end{figure}

\begin{figure}
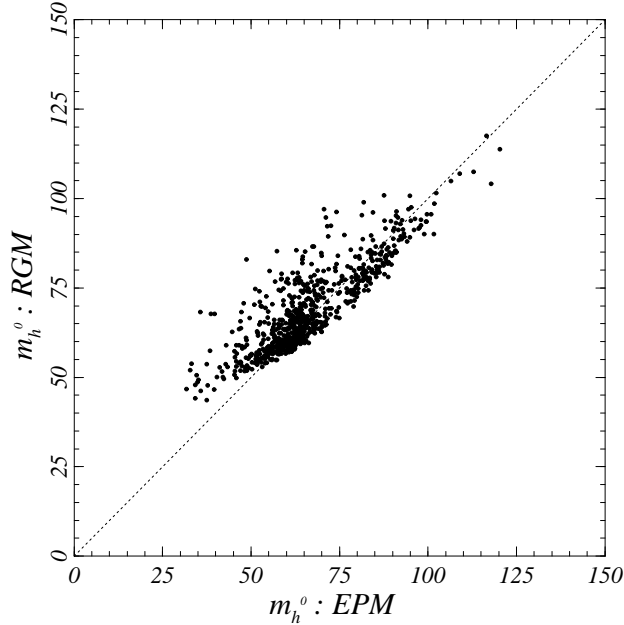

\caption{
Same as in Fig.\ 5,  except $Q = 600$ GeV
and the RGM calculation is carried out numerically.
}
\label{fig:f5}
\end{figure}

To further compare the two methods,
it is useful to consider the following exercise.
The RGM calculation could be modified
so that leading logarithms are not summed
to all orders,
e.g., one could use the one-loop leading-logarithm
approximation formula of Ref. \cite{haberham}.
Alternatively, we could minimize the unsummed
$\ln({m_{\tilde{t}}/Q})$ leading (and/or next-to-leading)
logarithms in the EPM by choosing
$Q \approx m_{\tilde{t}}$. In either case the residual
terms in the different calculations are qualitatively
brought on the same footing.
We carried out that exercise numerically, and
Fig.\ \ref{fig:f4}  (Fig.\ \ref{fig:f5})
compare the EPM with $Q = M_{Z}$ ($Q =600$ GeV) calculation
to the RGM leading logarithm approximation
(numerical calculation).
Once the residual two-loop
dependence of the calculation is properly adjusted,
the two algorithms agree well,
in particular for $Q \approx 600$ GeV.
[All calculations presented in Fig.\ \ref{fig:f4} (Fig.\ \ref{fig:f5})
use the $t$-quark
pole (running) mass.]

This all points at the most important source
of uncertainty: residual two-loop
(and wave-function) scale dependences
and a $\sim 5\%$ $\alpha_{s}$ dependence of $m_{t}$.
Two-loop terms fuel the ambiguity, which therefore cannot
be removed in a satisfactory fashion
within a one-loop
calculation. (Even though we could identify
some two-loop terms, the consistency of their removal
is doubtful).
Expanding the calculation to higher orders
is not straightforward. We discuss two-loop
calculations in Appendix \ref{sec:app4}.

Figs.\ \ref{fig:f2} and \ref{fig:f3} define the theoretical
ambiguity in determining
the loop induced $m_{h^{0}}$ in this scenario.
While the former sets the upper bound
(for a given $m_{t}^{pole}$), the latter is
a more accurate {\bf leading-logarithm} calculation
(and uses properly the running $t$-quark mass).
However,
within a one-loop calculation, and
in particular when next-to-leading logarithms and some finite
terms are comparable to higher-loop leading logarithms
(which is the case here -- see Appendix \ref{sec:app4}),
neither method provides a more consistent calculation
then the other.
As we argue in Appendix \ref{sec:app4}, the former
is also an upper bound on two-loop calculations,
which most probably diminish $m_{h^{0}}$.
We do not think that any stronger
statements are justified at present.
We continue to discuss the prediction and its
upper bound in the next section.

A different issue is that of the constraints
on the $t$-scalar left-right mixing enhancement
to $\Delta_{h^{0}}$. In the EPM [e.g., Eq. (\ref{mh})]
the enhancement is a straightforward result of diagonalizing
the CP-even mass matrix.
In the RGM one has to define the decoupling scale for the
$t$-scalars. If $m_{{\tilde{t}}_{1}} \gg  m_{{\tilde{t}}_{2}}$
then one would decouple only $\tilde{t_{1}}$ while ${\tilde{t}}_{2}$
loops could still  contribute to $\Delta_{h^{0}}$.
(In practice, the effect is accounted for
in the RGM by a set of appropriate
matching conditions.)
However, the amount of mixing $\hat{\Gamma}$ (and thus
the enhancement) is strongly constrained by
the physical vacuum.

Requiring that the SM extremum is a minimum
(i.e., no negative squared masses corresponding to physical
degrees of freedom), and furthermore, is the global minimum
(i.e., there is no deeper minimum that does not conserve
color and/or charge) strongly constrains $\hat{\Gamma}$.
Constraints imposed by the latter may
be evaded\footnote{This claim, however,
does not take into account finite-temperature effects.}
if the non-SM
global minimum is separated from the SM local minimum
by a tunneling time greater than the age of the universe
\cite{ccb5}.
Constraint (\ref{ccb3a}) (naively) attempts to eliminate
all, i.e., local and global, negative-energy CCB minima
in that region of parameter space and is too strong.
The points marked in the scatter plots
by only a an open ``diamond'' survive a more
careful analysis, i.e., they correspond to only
local minima\footnote{
In fact, all points are found to have at least a negative-energy
CCB local minimum of type $M^{2} = 0$ -- see
Appendix \ref{sec:app2}.
However, we find that in our region
of the parameter space
(\ref{ccb3a}) is a useful tool to identify
minima that
compete with the standard model minimum.}.
We find, using numerical methods, that
for $m_{t}^{pole} \lesssim 165$  GeV
(165 GeV $\lesssim m_{t}^{pole}$)
roughly
$80\%$ ($15\%$) of the points that fail to satisfy constraint
(\ref{ccb3a}) correspond to
(probably unacceptable) GCCB minima of the
(one-loop improved) tree-level scalar potential.
The enhancement is then partially washed away by requiring
consistency with the physical vacuum.
The possibilty of coexistence of the two vacua
(with a tunneling time greater than the age of the universe)
is, at the present stage of our
calculation, a remaining ambiguity.
We further discuss those issues in the next section and,
in greater detail, in  Appendix \ref{sec:app2}.

\section{The prediction for the Higgs boson mass
and its upper bound}
\label{sec:s5}

\begin{figure}
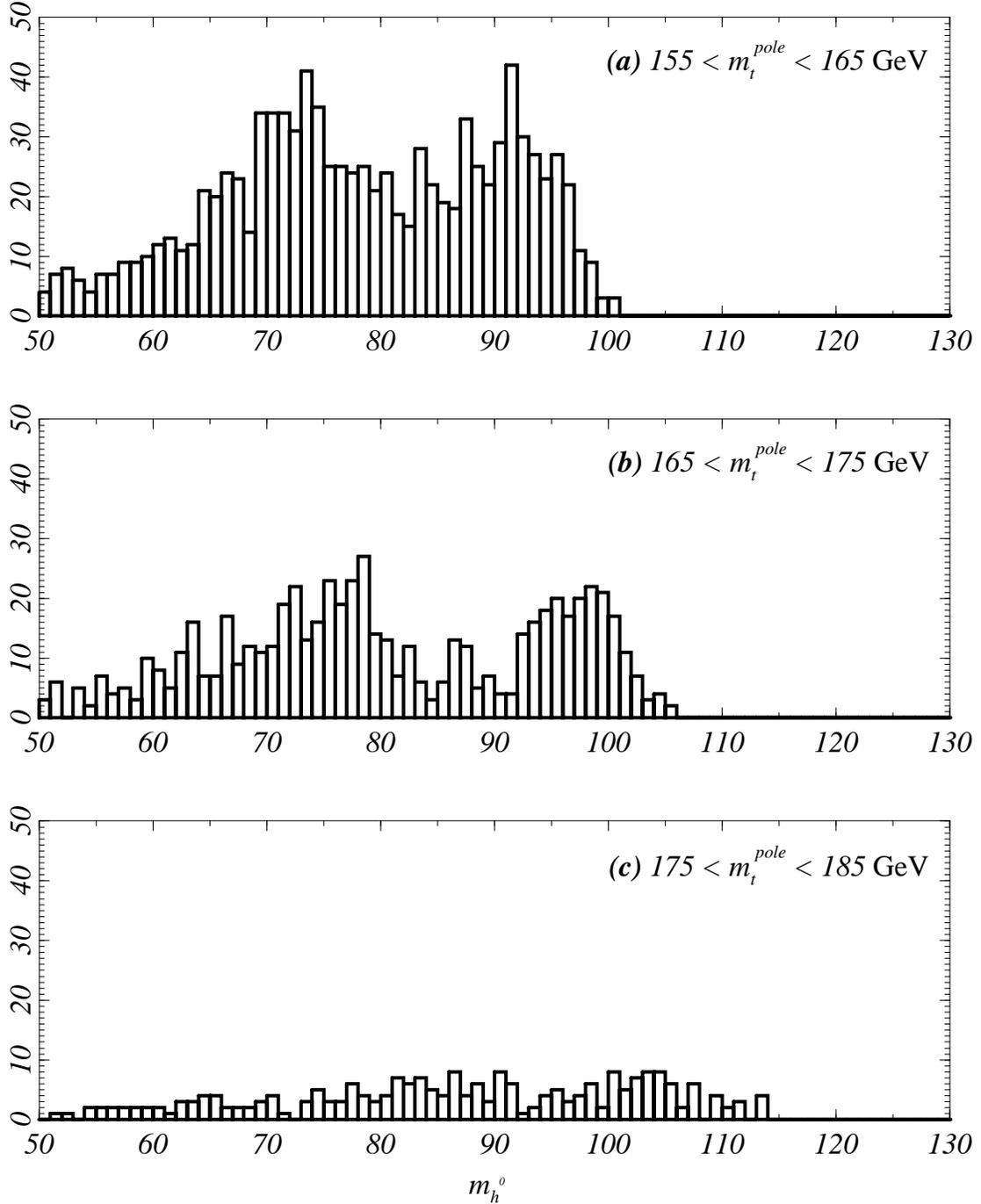

\caption{
The Higgs boson mass (in GeV) distribution
in a sample monte-carlo calculation
(using the EPM with the subtraction scale $Q$ at the $Z$-pole)
for the $t$-quark pole mass in the range
($a$) $[155,\,165]$, ($b$) $[165,\,175]$,
($c$) $[175,\,185]$  GeV.
}
\label{fig:f6}
\end{figure}

In this section we present and discuss
our numerical results. In section \ref{sec:s4}
we discussed various methods to calculate the mass (to one-loop).
The effective potential method (EPM) with a subtraction
scale $Q = M_{Z}$ (e.g., see Fig.\ \ref{fig:f2})
puts the weaker constraints on $m_{h^{0}}$ and is the conservative
choice for the upper bound.
In Fig.\ \ref{fig:f6}
we present monte-carlo distributions of $m_{h^{0}}$
using that method
for $155 \leq m_{t}^{pole} \leq 165$,
$165 \leq m_{t}^{pole} \leq 175$, and
$175 \leq m_{t}^{pole} \leq 185$ GeV.
The upper bound can be read off the figures,
i.e., $m_{h^{0}} \lesssim$ 101, 106, 114 GeV
for $m_{t}^{pole} \lesssim$ 165, 175, 185 GeV, respectively.
(Note that these bounds are deduced from our monte-carlo calculations,
and are not rigorous.)
Nevertheless, we would like to stress that typically $m_{h^{0}}$
lies below its upper bound.
Indeed, one cannot rule out the possibility
(if $m_{t}^{pole}$
is heavy enough) that $m_{h^{0}}$
is just outside the range that LEPII may cover.
However, this is not likely on both grounds,
the  $m_{t}^{pole}$ distribution [Eq. (\ref{mtop})] and
the $m_{h^{0}}$ (theoretical) distribution (as reflected in Fig.\
\ref{fig:f6}).

The role of the large left-right mixing in the $t$-scalar
sector can be seen in Figs.\
\ref{fig:f6}a and \ref{fig:f6}b, which exhibit
two peaks in the mass distribution,
a ``logarithmic'' and a ``mixing'' peak.
The former is due to the diagonal entries in the $t$-scalar mass
matrix [Eq. (\ref{stop})] while the latter, which is at larger mass,
is due to the mixing enhancement [$\Delta_{\theta_{t}}$
in Eq. (\ref{mixing})].
The interplay between the mixing and the $m_{t}^{4}$  factor
in the loop correction $\Delta_{h^{0}}$ [e.g., in Eq. (\ref{mh})],
and between
$\Delta_{h^{0}}$ and the tree-level mass $m_{h^{0}}^{T}$ (that grows
with $\tan\beta$ and, therefore, with $m_{t}^{pole}$)
can be seen by comparing Figs.\ \ref{fig:f6}a-b with
Fig.\ \ref{fig:f6}c.
The mixing enhancement plays an important role
for smaller values of
$m_{t}^{pole}$ ($\Delta\beta \rightarrow 0$).
In fact, that interplay generates for a fixed set of soft parameters
a local minimum in the $m_{h^{0}}$
distribution  (not shown in the figures)
near $m_{t}^{pole} \approx  165$ GeV.

\begin{figure}
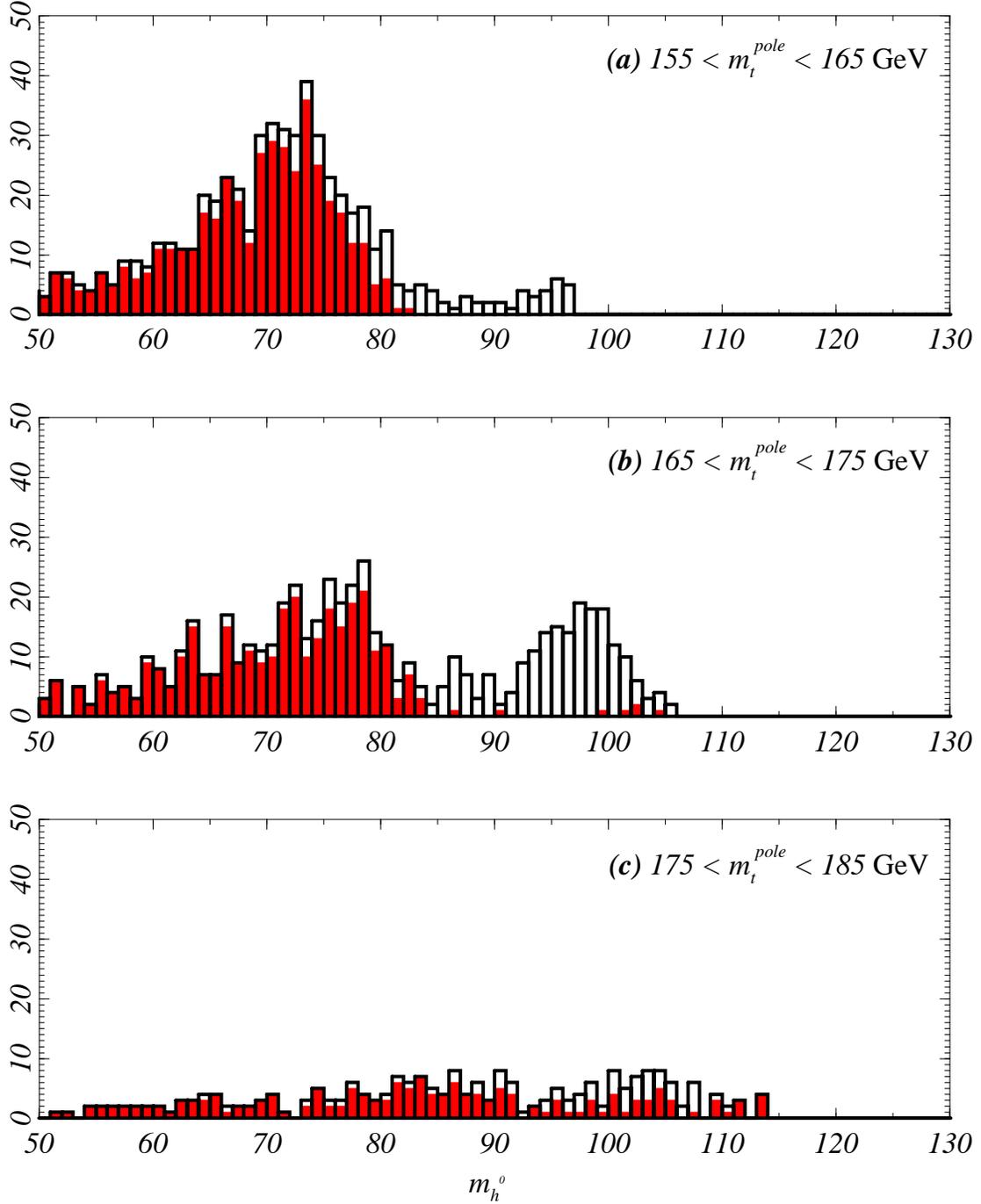

\caption{
Same as in Fig.\ 7 except for
the omission of points which correspond to a global
color and/or charge breaking minimum
(marked by a filled diamond in Fig.\ 3).
For comparison,
the shaded areas indicate points which are consistent
with the constraint (13)
(marked by a circle in Fig.\ 3).
The difference between the two distributions is
points with only a local CCB minimum
unnecessarily excluded by (13)
(marked by open diamonds in Fig.\ 3).
}
\label{fig:f7}
\end{figure}

The left-right mixing enhancement is strongly constrained when
requiring that the minimum of the full scalar potential is the physical one,
i.e., no fields other than $H_{1}^{0} = \nu_{down}$ and
$H_{2}^{0} = \nu_{up}$ have non-vanishing expectation values
(see section \ref{sec:s3} and Appendix \ref{sec:app2}).
Constraint (\ref{ccb3a}) eliminates potentially dangerous points in the
parameter space (marked by a ``diamond''
in Figs.\ \ref{fig:f2} and \ref{fig:f3}),
some of which, however, correspond to only a local
color-breaking minimum (marked by an open ``diamond'').
Those which correspond to
global minima are marked by a filled ``diamond''.
Fig.\ \ref{fig:f7} is the same as Fig.\ \ref{fig:f6}
except that points which
correspond to global color  and/or charge breaking
(GCCB) minima are omitted (the shaded areas
have no negative-energy CCB minimum, global or local.)
The mixing peak has diminished significantly
but has not disappeared.
Its presence is even stronger for larger values of $m_{t}^{pole}$
(see appendix \ref{sec:app2}).
(Some caution is required: the cosmological history
of the universe may accommodate such a global minimum.)

\begin{figure}
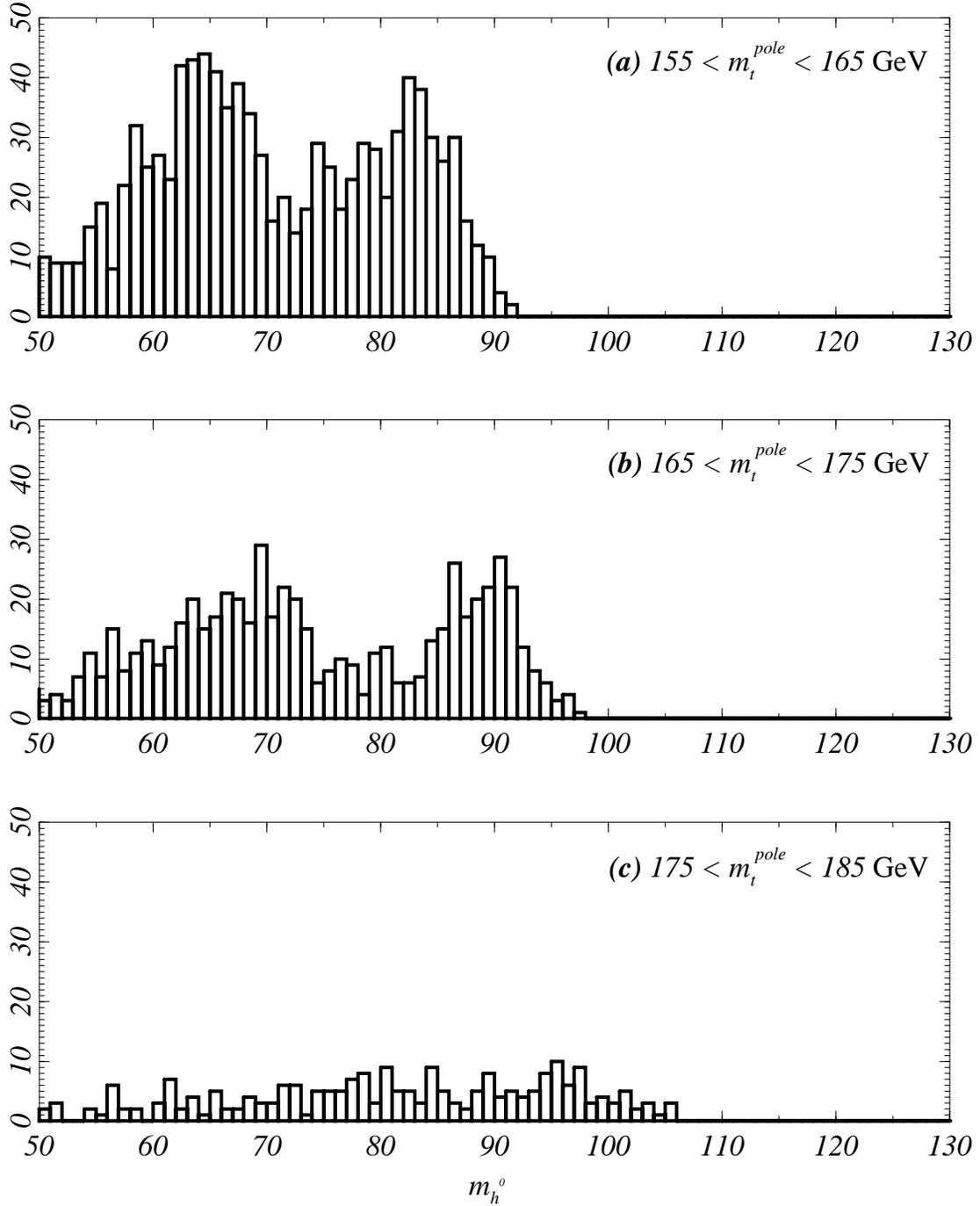

\caption{
Same as Fig.\ 7 except for substituting
the $t$-quark running (rather than pole) mass in the mass
expressions for the scalars.
}
\label{fig:f8}
\end{figure}

\begin{figure}
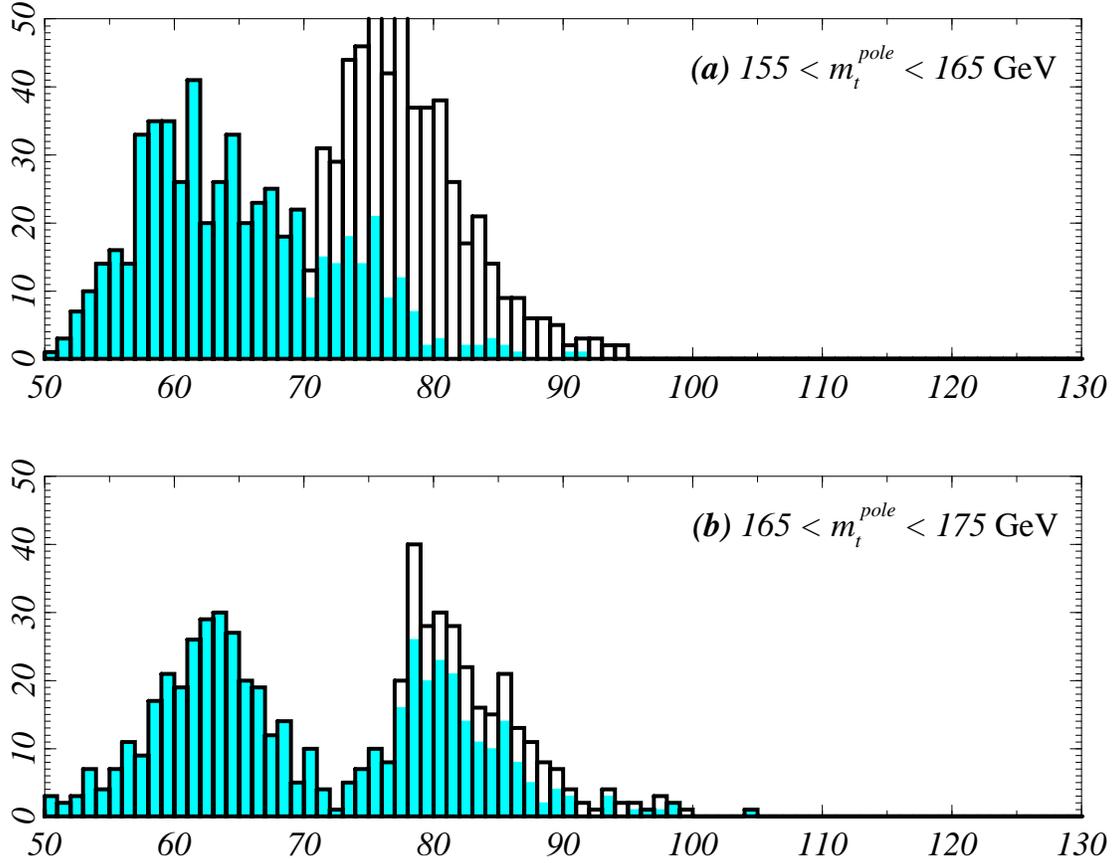

\caption{
Same as in Fig.\ 7a-b except
using the RGM (with numerical integration).
The distributions omitting
points which correspond to a global
color and/or charge breaking minimum
(marked by a filled diamond in Fig.\ 4)
(lightly shaded areas) are also indicated, and should be compared
with the total (shaded and unshaded)
areas in Fig.\ 8.  (For a $t$-quark heavier than about
180 GeV the 1HDM assumption that underlies the calculation
is not always accurate --
see Appendix C.)
}
\label{fig:f9}
\end{figure}

A different issue is that of the large two-loop ambiguity
in the calculation pointed out in section \ref{sec:s4}.
For example, in Fig.\ \ref{fig:f8} we show the same distributions
as in Fig.\ \ref{fig:f6}, but in which the running
(rather than the pole) $t$-quark mass was used
to calculate $m_{h^{0}}$.
(The running mass is $\sim 5\%$ smaller.)
Also, the distribution corresponding to the
renormalization group method (RGM) calculation
(e.g., see Fig.\ \ref{fig:f3}),
which we showed\footnote{
The RGM calculation sums the leading logarithms
to all orders in perturbation theory, while
in the EPM calculation with $Q \approx m_{\tilde{t}}$
higher order terms are, in principle, minimized,
and do not need to be summed.}
in Fig.\ \ref{fig:f5} to roughly agree with the EPM
calculation (with $Q \approx  m_{\tilde{t}}$),
is shown in Fig.\ \ref{fig:f9}
for $155 \leq m_{t}^{pole} \leq 165$ and
$165 \leq m_{t}^{pole} \leq 175$,
respectively.
Again, the $t$-quark running mass is used.
Also  shown are the distributions when GCCB
points are omitted (shaded areas)
to be compared with the total areas (shaded  and unshaded) in
Fig.\ \ref{fig:f7}.
Note the concentration
at lighter masses compared to  previous figures,
even though the upper limits are only slightly changed.
It should be stressed that
had we not included the proper RGM matching functions
\cite{ham} (which would roughly correspond to setting
the left-right mixings to zero by hand) then
the RGM distribution would sharply peak
at the lower end of the $m_{h^{0}}$ range.
Thus, the importance of the matching functions is obvious.

All, Figs.\ \ref{fig:f6} - \ref{fig:f9}, are consistent within
the theoretical ambiguity.
It is suggestive that a consistent
two-loop calculation will reduce the upper bounds given
by the EPM calculation (we choose conservatively Fig.\ \ref{fig:f6}),
but probably not to lower values than suggested by the
RGM (Fig.\ \ref{fig:f9}).
Further study is required before any stronger
conclusions can be drawn. Even so, for $m_{t}^{pole} \lesssim 175$ GeV
$m_{h^{0}}$ [in case $(1)$]
is within the reach of LEPII, and possibly of the Tevatron
(see Ref. \cite{htevatron}).
If $m_{t}^{pole}$ is lighter, as suggested by electroweak
precision data, our histograms imply that $m_{h^{0}}$
may still be relevant for Higgs searches at LEPI.
(In particular, if two-loop corrections
reduce $m_{h^{0}}$ as suggested by Fig.\ \ref{fig:f9}.)

\section{Conclusions}
\label{sec:s6}

To summarize, we worked in a constrained framework
assuming (up to high and low-scale threshold corrections)
$\alpha_{1}=\alpha_{2}=\alpha_{3}=\alpha_{G}$,
$h_{b} = h_{\tau}$,
and universal initial conditions.
Also, we explicitly included the correlation between
the $t$-quark mass and the weak angle
from electroweak precision data.
Within that framework, only $\tan\beta \approx 1$
and large $\tan\beta$ regions are allowed.
In the large $\tan\beta$ region (which is allowed
only for a large $m_{t}^{pole}$, unless $m_{b}$
is allowed to get large finite corrections
from superpartner loops)
$m_{h^{0}}$ is bounded by the triviality bound, and
is probably not interesting for Higgs searches at LEPII.

In the $\tan\beta \approx 1$ region
the tree-level
$m_{h^{0}}^{T}$ nearly vanishes, but grows with $m_{t}^{pole}$.
The loop correction to the mass is enhanced by
large $\tilde{t}_{L} - \tilde{t}_{R}$
mixing terms.
The latter are diminished with an increasing $m_{t}^{pole}$
but, as $m_{t}^{pole}$ grows, the overall factor
of $m_{t}^{4}$ can compensate for that effect.
The enhancement due to the large mixing is
further constrained
by requiring the physical vacuum to correspond to
the global minimum of the full scalar potential.
We find that for parameters corresponding to large
mixing there is usually a color breaking minimum,
but in many cases this is only a (presumably harmless)
local minimum.
We also pointed out that a two-loop ambiguity in the calculation
of the Higgs boson mass is enhanced in our case
and can be as large as 20 GeV.
All these effects lead to the conservative bound
$m_{h^{0}}\lesssim 100$ (110) GeV for $m_{t}^{pole} \lesssim 160$
(175) GeV.
(We briefly discuss two-loop calculations in Appendix \ref{sec:app4},
where we conclude that such a calculation will most probably
strengthen the upper bound.)
The Higgs boson mass in the class of models characterized by
$\tan\beta \approx 1$
is then almost certainly  within the reach of LEPII
(and possibly within the reach of the Tevatron \cite{htevatron}).
Furthermore, the Higgs boson
is typically lighter than the upper bound, and
may be light enough to still be discovered at LEPI.

Regarding the $t$-quark mass,
one requires for $SU(2)\times U(1)$ breaking (in this scenario)
$m_{t}^{pole} \gtrsim 140$ GeV, and further requiring
$\tan\beta \gtrsim 1.1$ (which is the case if the divergent
limit is to be avoided, i.e., if the divergence
is not stabilized by model-dependent finite terms)
we obtain $m_{t}^{pole} \gtrsim 155$ GeV.
Eq. (\ref{mtop}) and, in particular, the structure of the
radiative corrections to the $Z \rightarrow b\bar{b}$ vertex,
however, suggest that $m_{t}^{pole}$ is not much larger.
(The latter can be evaded only
if both the  $t$-scalar and the Higgsino
are light, which is not very likely in our case.)
Also, we were able to understand the smallness of the
parameter space allowed by $h_{b} = h_{\tau}$ in
terms of a custodial symmetry. The symmetry (theoretical)
argument suggests $m_{t}^{pole} \lesssim 185$ GeV
($\Delta_{\beta} \lesssim 1$).
In Appendix \ref{sec:app3} we briefly examine extended
Higgs sectors, and point out that the Higgs mass is a possible
probe of such extensions if $m_{t}^{pole}$ (and/or $\tan\beta$)
is known.

In Appendix \ref{sec:app2} we point out that the full (one-loop
improved) scalar potential generically possesses color
and/or charge breaking negative-energy local minima.
Analytic constraints that are often found in the literature
attempt to eliminate all such local minima
(i.e., are too strong) but by construction
apply to only certain types of minima while ignoring others
(i.e., are too weak).
We give a general classification
of the local minima and are able to further identify [in case $(1)$]
those that can potentially compete with the standard model minimum.

\acknowledgments
This work was supported by the US Department of Energy
Grant No. DE-AC02-76-ERO-3071. It is pleasure to thank
J. Erler for useful discussions
concerning electroweak observeables, and to A. Pomarol
for useful discussions and for his comments on the manuscript.

\appendix
\section{The Higgs sector custodial symmetries}
\label{sec:app1}

In this appendix we examine some symmetry aspects of the SGUT solutions,
and in particular, of case $(1)$.
Some features of the latter, as well as the distinction between the
two cases, can be better understood in terms of the custodial
$SU(2)_{L + R}$ approximate symmetry of the SM lagrangian \cite{cus};
i.e., turning off hypercharge
and flavor mixings,
and if $h_{t} = h_{b} = h$, then one can rewrite
the $t$ and $b$ Yukawa terms as an
$SU(2)_{L}\times SU(2)_{R}$ invariant, i.e.,
\begin{equation}
h\left(\begin{array}{c}t_{L} \\ b_{L} \end{array} \right)_{a}
\epsilon_{ab}
\left(\begin{array}{cc}H_{1}^{0} & H_{2}^{+} \\
H_{1}^{-} & H_{2}^{0} \end{array} \right)_{bc}
\left(\begin{array}{c}-b^{c}_{L} \\ t^{c}_{L} \end{array} \right)_{c}
\label{matrixeq}
\end{equation}
(where in the SM $H_{2} = i\tau_{2}H_{1}^{\ast}$)
and $SU(2)_{L}\times SU(2)_{R} \rightarrow SU(2)_{L+R}$
for $\nu \neq 0$.
However, $h_{t} \neq h_{b}$ and the different hypercharges of
$t^{c}_{L}$ and $b^{c}_{L}$ explicitly break
the left-right symmetry, and therefore the residual custodial
symmetry.

In the MSSM, on the other hand,  $H_{1}$ is distinct from $H_{2}$
and if $\nu_{down} \neq \nu_{up}$
$SU(2)_{L}\times SU(2)_{R} \rightarrow U(1)_{T_{3L} + T_{3R}}$
\cite{cus2}. In particular, the
$SU(2)_{L + R}$ symmetry
is preserved if $\beta = \frac{\pi}{4}$ but is
maximally broken if $\beta = \frac{\pi}{2}$.
Ignoring hypercharge, we can then define
for illustration two quantities,
\begin{mathletters}
\label{breaking}
\begin{equation}
B_{M} \equiv \nu_{down}\frac{h_{t} - h_{b}}{m_{t} - m_{b}},
\label{breaking1}
\end{equation}
\begin{equation}
B_{H} \equiv m_{b}\frac{\tan\beta - 1}{m_{t} - m_{b}},
\label{breaking2}
\end{equation}
\end{mathletters}
that  measure the $SU(2)_{L + R}$
symmetry breaking in the matter
and Higgs sectors, respectively.
If we map the $\tan\beta - m_{t}^{pole}$ plane onto the
$B_{M} - B_{H}$ plane, we find that cases $(1)$ and $(2)$
approximately correspond to $(1,0)$ and $(0,1)$, respectively.
Thus, while the constrained $SO(10)$ solution realizes
$m_{t} \neq m_{b}$ as an intrinsic property
of the vacuum (i.e., $\nu_{up} \gg \nu_{down}$) and the grand-unified
symmetry protects $B_{M} \approx 0$ \cite{warsaw},
the $\tan\beta \approx 1$ case realizes  the $SU(2)_{L+R}$
(explicit) breaking as a scaling phenomenon (i.e., $B_{M} \propto
h_{t}^{fixed}$)
and the vacuum respects the symmetry (up to loop corrections).

The symmetry that is spontaneously broken
in case $(2)$ is a $O_{4} \times O_{4}$ symmetry.
Haber and Pomarol observed \cite{cus3}
that for $m_{3} \rightarrow 0$ [which is the situation in case $(2)$]
there is no mixing
between $H_{1}$ and $H_{2}$ and
the Higgs sector respects $O_{4} \times O_{4}$
(up to gauge-coupling corrections).
The symmetry is broken to $O_{3} \times O_{3}$
for $\nu_{down} \neq \nu_{up} \neq 0$ and the six Goldstone bosons
are the three SM Goldstone bosons, $A^{0}$, and $H^{\pm}$.
The symmetry is explicitly broken
for $g_{2} \neq 0$ (so $m_{H^{+}} = M_{W}$)
and is not exact
even when neglecting gauge couplings
(i.e., $m_{3} \neq 0$).
Thus, $A^{0}$ and $H^{\pm}$ are massive pseudo-Goldstone bosons,
i.e.,
$m_{H^{+}}^{2} - M_{W}^{2} \approx m_{A^{0}}^{2} = C\times m_{3}^{2}$.
However, $C = -\frac{2}{\sin2\beta}$ and can be large,
which is a manifestation of the fact
that
$O_{4} \times O_{4} \rightarrow O_{4} \times O_{3}$
for $\nu_{down} = 0$.
(The limit $m_{3} \rightarrow 0$ corresponds also to a $U(1)$
Peccei-Quinn symmetry \cite{hall}.)

Our main interest in this work is, however, case $(1)$, where
from (\ref{min2})
the Higgs-sector $SU(2)$ custodial symmetry constrains
\begin{equation}
m_{H_{1}}^{2} + \mu^{2} = m_{H_{2}}^{2} + \mu^{2} = -m_{3}^{2}.
\label{cusH}
\end{equation}
This is  realized by taking the limit $|\mu| \rightarrow \infty$,
provided $|B| \approx |\mu|$ which in our
formalism comes out as a prediction.
(Indeed, many SGUT models can easily realize a large $B_{0}$,
for example, see Ref. \cite{b}.)
Relation (\ref{cusH}) results in a (tree-level) massless
eigenvalue in the neutral CP-even mass matrix. (Note that the
SM Goldstone bosons are also massless Goldstone bosons
of the broken $SU(2)\times SU(2)$.)

The smallness of $\Delta_{\beta}$ is protected
by the custodial symmetry, i.e., $\Delta_{\beta} \neq 0$
is explained by loop corrections.
The broken symmetry in the Yukawa sector induces at the loop level
a small breaking in the Higgs sector
(e.g., $m_{H_{1}}^{2} \neq m_{H_{2}}^{2}$). The magnitude
of the effect can be estimated by integrating
$d\tan\beta/\tan\beta$, i.e.,
\begin{equation}
\Delta_{\beta}(M_{Z}) = \Delta_{\beta}(M_{G})
- \frac{3}{16\pi^{2}}\int_{\ln\frac{M_{G}}{M_{Z}}}^{0}(h_{t}^{2} - h_{b}^{2})
d\ln\frac{Q}{M_{Z}} \approx \Delta_{\beta}(M_{G}) + 0.6,
\label{tanb}
\end{equation}
where we neglected $h_{\tau}$, as well as threshold corrections,
took $h_{t} \approx 1$, $h_{b} \approx 0$, and
assumed $\Delta_{\beta} < 1$.
Thus, even though the parameter space in case $(1)$ is severely
restricted, we can understand the restriction in terms of a symmetry.

\section{Constraints on color and charge breaking minima}
\label{sec:app2}

Let us consider the super-potential
\begin{equation}
W = h_{t}{\tilde{t}}_{L}{\tilde{t}}_{R}H_{2}^{0} + \mu H_{1}^{0}H_{2}^{0},
\label{w}
\end{equation}
where we have performed an $SU(2)$ rotation
so that $H_{2}^{+}$ has no vacuum expectation value.
(We do not distinguish
our notation for a superfield from that of its scalar component.)
It is easy to convince oneself that only those terms
in the super-potential
are relevant when searching for the global minimum
of the scalar potential in the large
$h_{t}$ limit \cite{ccb2}.
Using standard techniques we arrive at the corresponding
scalar potential:
\begin{equation}
h_{t}^{2}V = M^{2} - \Gamma + \Lambda,
\label{pot1}
\end{equation}
where the bilinear, trilinear, and quartic terms are
\begin{equation}
M^{2} = m_{1}^{2}{H_{1}^{0}}^{2} +
m_{2}^{2}{H_{2}^{0}}^{2} +
2m_{3}^{2}H_{1}^{0}H_{2}^{0} +
m_{{\tilde{t}}_{L}}^{2}{\tilde{t}}_{L}^{2} +
m_{{\tilde{t}}_{R}}^{2}{\tilde{t}}_{R}^{2},
\label{bigm}
\end{equation}
\begin{equation}
\Gamma = \left| 2\left(|A_{t}|H_{2}^{0} + s|\mu|H_{1}^{0}\right)
{\tilde{t}}_{L}{\tilde{t}}_{R}\right|,
\label{gamma}
\end{equation}
and
\begin{equation}
\Lambda = ({\tilde{t}}_{L}^{2} + {\tilde{t}}_{R}^{2}){H_{2}^{0}}^{2}
+ {\tilde{t}}_{L}^{2}{\tilde{t}}_{R}^{2}
+ \frac{1}{h_{t}^{2}}(\frac{\pi}{2}\times``D-{\mbox{terms''}}),
\label{lambda}
\end{equation}
respectively.
All fields were scaled $\phi \rightarrow \phi/h_{t}$ and are
taken to be real and positive (our phase
choice for the fields, which fixed $m_{3}^{2} < 0$
and $\Gamma > 0$, i.e., maximized the negative
contributions to $V$)
and all parameters are real.
$m_{1,\,2}^{2} = m_{H_{1,\,2}}^{2} + \mu^{2}$,
$s = \mu A_{t}/|\mu A_{t}|$, and the expression for the $``D-$terms''
is \cite{ccb2}
\begin{eqnarray}
\frac{12}{5}\alpha_{1}
\left[-\frac{{H_{1}^{0}}^{2}}
{2} + \frac{{H_{2}^{0}}^{2}}{2} + \frac{{\tilde{t}}_{L}^{2}}{6}
- \frac{2{\tilde{t}}_{R}^{2}}{3}\right]^{2}
+ \alpha_{2}\left[{H_{1}^{0}}^{2}
- {H_{2}^{0}}^{2} + {\tilde{t}}_{L}^{2}\right]^{2}
+ \frac{4}{3}\alpha_{3}\left[{\tilde{t}}_{L}^{2}
- {\tilde{t}}_{R}^{2}\right]^{2}.  & &
\label{dterms}
\end{eqnarray}
We can parametrize the four fields in terms
of an overall scale $X$ and three angles
$0 \leq \alpha,\,\beta,\, \gamma \leq \frac{\pi}{2}$:
$H_{1}^{0} = X\sin{\alpha}\cos{\beta}$,
$H_{2}^{0} = X\sin{\alpha}\sin{\beta}$,
${\tilde{t}}_{R} = X\cos{\alpha}\cos{\gamma}$,
${\tilde{t}}_{L} = X\cos{\alpha}\sin{\gamma}$,
and redefine
\begin{equation}
h_{t}^{2}V(X) = M^{2}(\alpha , \,\beta , \,\gamma)X^{2} -
\Gamma(\alpha , \,\beta , \,\gamma)X^{3} +
\Lambda(\alpha , \,\beta , \,\gamma)X^{4}.
\label{pot2}
\end{equation}
Then, for fixed angles, $V(X)$ will have a minimum
for $X \neq 0$ provided the condition
$32M^{2}\Lambda < 9\Gamma^{2}$ is satisfied.
In that case,
\begin{equation}
X_{min} = \frac{3}{8}\frac{\Gamma}{\Lambda}\left[1 +
\left(1 - \frac{32M^{2}\Lambda}{9\Gamma^{2}}\right)^{\frac{1}{2}}
\right] \geq 0,
\label{xmin}
\end{equation}
and
\begin{equation}
h_{t}^{2}V_{min} = -\frac{1}{2}X_{min}^{2}\left(
\frac{\Gamma}{2}X_{min} - M^{2} \right).
\label{pot3}
\end{equation}

We only allow parameters for which a standard model (SM) minimum
exists.
The SM minimum corresponds to $\alpha = \frac{\pi}{2}$ and
$\beta = \beta^{0}$ ($\gamma$ is irrelevant
and $\tan\beta^{0} = \nu_{up}/\nu_{down}$ is the angle
used to fix $\mu$, $m_{3}$, as well as the Yukawa couplings).
It is easy to convince
oneself that in that limit the $4 \times 4$ second-derivative
matrix is $2 \times 2$ block diagonal
(otherwise baryon number is violated).
Thus, it is sufficient to confirm that the four physical
eigenvalues are positive to ensure that
it is a minimum. (This is done in our numerical
calculations.)
If these conditions are satisfied then the SM is at least
a local (negative-energy) minimum, and one has
$\Gamma_{\mbox{\tiny SM}} = 0$,
$M^{2}_{\mbox{\tiny SM}} < 0$, and (\ref{xmin})
and (\ref{pot3}) reduce to the usual results
$X_{min}^{\mbox{\tiny SM}} = \sqrt{- M^{2}_{\mbox{\tiny SM}}/2
\Lambda_{\mbox{\tiny SM}}}$,
$h_{t}^{2}V_{min}^{\mbox{\tiny SM}}=
- M^{4}_{\mbox{\tiny SM}}/4\Lambda_{\mbox{\tiny SM}}$.

Let us now consider the possibility
of additional CCB minima with $\cos\alpha \neq 0$.
Since $m_{\tilde{t}_{L,\,R}}^{2} > 0$
for the class of models we are considering this
requires $\Gamma \neq 0$, which we assume hereafter.
{}From (\ref{pot2}) -- (\ref{pot3})
it is easy to classify the possible color
and/or charge breaking (CCB) minima for definite
$\alpha$, $\beta$, $\gamma$.
One finds
that for $\Gamma^{2} \leq 32\Lambda M^{2}/9$
there is no CCB minimum, while for
$32\Lambda M^{2}/9 < \Gamma^{2} \leq 4\Lambda M^{2}$
the CCB minimum exists but has
$V_{min}^{\mbox{\tiny CCB}} > 0 > V_{min}^{\mbox{\tiny SM}} $,
which is presumably safe.
For $4\Lambda M^{2} < \Gamma^{2}$
(including the more rare case
$M^{2} < 0$, that must fall in this category)
there is a negative-value CCB minimum, which may however be
either local (presumably safe), i.e.,
$V_{min}^{\mbox{\tiny CCB}} > V_{min}^{\mbox{\tiny SM}}$,
or global (probably  unacceptable), i.e.,
$V_{min}^{\mbox{\tiny CCB}} < V_{min}^{\mbox{\tiny SM}}$.
Here, a sufficient (but not necessary)
condition for an acceptable model is
\begin{equation}
\Gamma^{2} \leq 4\Lambda M^{2},
\label{master}
\end{equation}
which generalizes that of Ref. \cite{ccb2}.
In principle, the above discussion holds for any number
of fields (i.e., any number of angles),
only the explicit expressions for $M^{2}$, $\Gamma$,
and $\Lambda$ are more complicated.
If constraint (\ref{master})
holds for every choice of $\alpha$, $\beta$, and $\gamma$
($\cos\alpha \neq 0$)
then there is no negative-valued
color and/or charge breaking minimum,
global (GCCB) or local.
In the special case $M^{2} < 0$ the constraint cannot be satisfied
and there will be a negative-energy local CCB minimum,
an observation made previously by Gunion et al. \cite{ccb2}.

If (\ref{master}) does not hold, further investigation
is needed to determine whether the minimum is global
or local. It is straightforward to show that (\ref{master})
is satisfied for $\sin\beta = 0$, i.e., there is no negative-valued
CCB minimum for $H_{2}^{0} = 0$ for the class
of models we are considering. We will therefore
restrict our attention to the case $H_{2}^{0} \neq 0$,
in which case it is convenient to reparametrize
$X = H_{2}^{0}$
and rescale all fields by an additional
factor of $1/H_{2}^{0}$.
This simplifies the discussion and enables us to examine
and generalize previous work on the subject
\cite{ccb1,ccb3,ccb4,ccb2}.

By inspection (\ref{master})
cannot hold for all values of the fields
and there is always  (and not just for $M^{2} < 0$, which is hard
to achieve in dangerous directions)
a local CCB minimum,
e.g., for $M^{2} = 0$ ($\Lambda$ is positive definite).
For example, one can choose
\begin{mathletters}
\label{ccb2}
\begin{equation}
H_{1}^{0} = -\frac{m_{3}^{2}}{m_{1}^{2}},
\end{equation}
\begin{equation}
{\tilde{t}}_{L}
= \frac{1}{m_{{\tilde{t}}_{L}}}\sqrt{
\frac{1}{2}\left( \frac{m_{3}^{4}}{m_{1}^{2}}
- m_{2}^{2}\right)},
\end{equation}
\begin{equation}
{\tilde{t}}_{R} = {\tilde{t}}_{L}
\frac{m_{{\tilde{t}}_{L}}}{m_{{\tilde{t}}_{R}}},
\end{equation}
\end{mathletters}
which is always possible because of (\ref{cond}).
We confirmed this in our numerical studies,
i.e., there is a CCB (negative-valued)
local minimum even if $V^{\mbox{\tiny SM}}
= V(\nu_{down},\,\nu_{up})$
is the global minimum.
(We do not consider additive constants.)
Although such local minima may be of cosmological interest,
the relevant question in our case is
whether $V^{\mbox{\tiny SM}}$
is the global minimum of the full (or more precisely, of the four-field)
scalar potential.
Similarly, we assume that positive-valued
CCB minima would not be populated and are therefore harmless.
The answer has to be given by numerical
mapping of all minima,
but it is still useful to review how one could derive analytic
constraints from (\ref{master}), which are typically
relevant only for specific regions of the parameter space.

For example, if we fix $H_{1}^{0} = 0$, $\tilde{t}_{L}=\tilde{t}_{R}=1$,
then (\ref{master}) gives the well known result
\cite{ccb1}
\begin{equation}
A_{t}^{2} \leq 3(m_{\tilde{t}_{L}}^{2} +m_{\tilde{t}_{R}}^{2}
+ m_{H_{2}}^{2} + \mu^{2}).
\label{ccb3}
\end{equation}
The equal field direction was chosen so the positive $D-$term
contribution $\propto 1/h_{t}^{2}$ vanishes. This is not a relevant
requirement for $h_{t} \approx 1$
(i.e., there may exist a deeper minimum
with non vanishing $D-$terms).
It is obvious that (\ref{ccb3}) is not relevant when
$|\mu| \rightarrow \infty$.
We find that a more useful constraint
is [taking  $H_{1}^{0} \approx 1$, $\tilde{t}_{L} \approx \tilde{t}_{R}
 \equiv t \ll 1$
and using (\ref{master})]
\begin{equation}
\left( |A_{t}| + s|\mu| \right)^{2} \leq
2(m_{\tilde{t}_{L}}^{2} +m_{\tilde{t}_{R}}^{2}).
\label{ccb4}
\end{equation}
Note that if the order $t^{2}$ corrections
to (\ref{ccb4}) are not negligible then $V$ is less negative,
which motivated our choice $t \ll 1$, i.e.,
the more dangerous direction.
That constraint was also used
by Drees et al. \cite{ccb3} to describe their numerical
results (i.e., if we take $m_{1} = m_{2}$ in their formula).
We confirmed in our numerical studies that (\ref{ccb4})
and not (\ref{ccb3}) is relevant in our case.
[We found nearly all points
to be consistent with (\ref{ccb3})
for our ranges of the soft parameters.]

Finally, we perform numerical minimization of
the potential (\ref{pot1})
and map all minima using monte-carlo
routines. We find that (\ref{ccb4}) is
to a good approximation sufficient
to avoid a GCCB minimum, but is not necessary.
For example, only about $80\%$ ($15\%$)
of the points which are inconsistent with
(\ref{ccb4}) correspond to GCCB
for $m_{t}^{pole} \lesssim 165$ GeV
($m_{t}^{pole} \gtrsim 165$ GeV).
We compare the points consistent
with (\ref{ccb4}) and with the numerical search
in Fig.\ \ref{fig:f7}.
The greater relevance of (\ref{ccb4}) for
$m_{t}^{pole} \lesssim 165$ GeV can be understood
if we recall that the lower
bound on $|\cos 2\beta|$ increases with $m_{t}^{pole}$,
and thus $V^{\mbox{\tiny SM}}$ becomes a deeper minimum.
(The lower bound is related
to $h_{t}$ unitarity, and that statement
is independent of our $h_{b} = h_{\tau}$  assumption.)
On the other hand, $|\mu|$
and the negative terms in $V^{\mbox{\tiny CCB}}$
[e.g., Eq. (\ref{pot1})] diminish
with increasing $m_{t}^{pole}$.
That interplay shifts the CCB minimum from a
global to a local minimum.

Let us demonstrate that claim in a simple-minded example assuming
$h_{t} \approx h_{t}^{fixed} \approx 1 - 1.1$.
We define ${\sin}^{2}\beta = 2m_{t}^{2}/h_{t}^{2}\nu^{2} \equiv x$,
$0.5 \leq x \leq 1$
[$\nu = 2M_{W}/g_{2}$].
We can rewrite
\begin{equation}
h_{t}^{2}V^{\mbox{\tiny SM}} = - \frac{1}{8}M_{Z}^{2}
h_{t}^{2}\nu^{2}\left[1 - 4x(1-x)\right].
\label{examp1}
\end{equation}
$\left| h_{t}^{2}V^{\mbox{\tiny SM}}\right|$ sharply increases
with $m_{t}^{pole}$ (i.e., with $x$).
On the other hand, taking
$m_{H_{2}}^{2} \approx M_{Z}^{2} \approx 0$ in (\ref{min1})
we have
\begin{equation}
\mu^{2} \approx m_{H_{1}}^{2}\left(\frac{1 - x}{2x - 1}\right),
\label{examp2}
\end{equation}
and $\mu^{2}$ decreases from $+\infty$ to 0 in
the above range of $x$.
Now assume that constraint (\ref{ccb4})
is not satisfied, i.e.,
\begin{equation}
h_{t}^{2}V^{\mbox{\tiny CCB}} \approx -\frac{1}{2}
\kappa^{2}\mu^{2}\left(\kappa\mu^{2} - m_{{\tilde{t}}_{L}}^{2}
- m_{{\tilde{t}}_{R}}^{2}\right)t^{2},
\label{examp3}
\end{equation}
where $t^{4} \ll t^{2}$ and $\kappa$ is a positive number
in the range $[\frac{3}{8},  \, \frac{3}{4}]$
($X_{min} = \kappa|\mu|$).
We also assumed $|A_{t}| \ll |\mu|$,
and neglected $t^{4}$ terms. ($M^{2} < 0$ would increase $\kappa$.)
To further simplify, consider the case $m_{0} = 0$, so that
$m_{{\tilde{t}}_{L}}^{2} \approx m_{{\tilde{t}}_{R}}^{2}
\approx 6 M_{\frac{1}{2}}^{2} \approx 12m_{H_{1}}^{2} $.
$V^{\mbox{\tiny CCB}} \lesssim V^{\mbox{\tiny SM}}$
in this case if
\begin{equation}
\frac{1}{4\kappa^{2}}\frac{M_{Z}^{2}h_{t}^{2}\nu^{2}}
{m_{H_{1}}^{4}t^{2}} \lesssim \left(\frac{1-x}{2x-1}\right)
\frac{\left[\kappa\left(\frac{1-x}{2x-1}\right)
- 24\right]}{1 - 4x(1-x)}.
\label{examp5}
\end{equation}
The $r.h.s.$ has to be positive, i.e., $x \rightarrow 0.5^{+}$
or $m_{t}^{pole} \lesssim 150$ GeV.
Slightly increasing $m_{t}^{pole}$ would reverse
the inequality.

Lastly, there are two caveats.
The first is that we used
in this section the (one-loop
improved) tree-level potential.
However, following Ref. \cite{grz} we perform the calculations
at the $t$-scalar scale
to eliminate large loop corrections.
Secondly, a GCCB minimum may be ``safe'' if separated from the local
standard-model minimum by a tunneling time greater
than the age of the universe \cite{ccb5}.
Such considerations are beyond the scope
of our present work and would also require a
consideration of finite-temperature effects.

\section{Calculation of the loop-induced mass}
\label{sec:app4}

\subsection{The EPM: Run and diagonalize}

A straightforward way to calculate $\Delta_{h^{0}}$ is
$(a)$ to take
the  first and second derivatives of $\Delta V(Q)$ with respect to
the neutral CP-even and CP-odd components of $H_{i}$ at the minimum;
$(b)$ to absorb the latter
in $m_{A^{0}}$; and $(c)$ to calculate the correction to (\ref{tree}):
the effective potential method (EPM).
Like Ref. \cite{tam1}, we will follow Ref. \cite{ez}, however,
we add to their expressions
``D-term'' \cite{brig} and Higgs-Higgsino and gauge-gaugino
\cite{haberdiaz} contributions, which are typically
$-(1-2)$ and $-(2-4)$ GeV, respectively.
Note that $(i)$ all the parameters are taken at the subtraction scale $Q$.
The mass matrices are thus calculated at the scale $Q$ and
$(ii)$ only then are diagonalized, i.e.,
a ``run and diagonalize'' algorithm.
Our two choices in section \ref{sec:s4} were $Q = M_{Z}$ and
$Q = 600$ GeV $\sim m_{\tilde{t}}$.
(Large
$\ln\frac{m_{A^{0}}}{M_{Z}}$ logarithms are multiplied by small
couplings and do not invalidate the loop expansion, i.e., the
EPM. This is equivalent to the statement above that
Higgs-Higgsino contributions are small.)
One corrects for threshold effects
by the subtraction of one-loop leading logarithms
(included in $\Delta V$), and thus to that order the two
choices are equivalent.

The $Q$ dependence of $m_{h^{0}}(Q)$ comes about from
$(i)$ logarithmic wave-function dependence of $\nu_{i}(Q)$ and
$(ii)$ two-loop implicit field-independent dependences of
$\Delta V(Q)  = \Delta V[m_{i}(Q),\,\alpha_{i}(Q),\,h_{t}(Q)]$
(where $m_{i}$ runs over the relevant mass parameters).
Both are often underestimated. The importance of $(i)$ is obvious, e.g.,
from the $m_{t}^{4}/M_Z^{2}$ factor. The field-independent dependence  $(ii)$
was studied in Ref. \cite{tam2}, where it was shown
to be of the order of two-loop (next-to-leading) logarithms.
One can instead examine directly the expression for $\Delta_{h^{0}}$, e.g.,
the leading logarithm in (\ref{mh}) evolves with scale (neglecting
$\Delta_{\theta_{t}}$ and multiplicative factors) roughly as
\begin{eqnarray}
&\frac{1}{8\pi^{2}}\ln\left(\frac{m_{\tilde{t}}^{2}(Q^{2})}{m_{t}^{2}}
\right) \approx
\frac{1}{8\pi^{2}}\ln\left(\frac{m_{\tilde{t}}^{2}(Q_{0}^{2})}{m_{t}^{2}}
\right) +\frac{1}{128\pi^{4}}
\left\{ \kappa h_{t}^{2}\left[\frac{m_{H_{2}}^{2} + A_{t}^{2}}
{m_{\tilde{t}}^{2}} + 2\right]
-\frac{16}{3}g_{s}^{2}\frac{M_{\tilde{g}}^{2}}{m_{\tilde{t}}^{2}}
\right\} \ln\frac{Q^{2}}{Q_{0}^{2}}, &
\label{q}
\end{eqnarray}
where $\kappa = 1,2$ for $\tilde{t}_{L,\,R}$, respectively,
and we took $m_{t}(Q) \approx m_{t}(Q_{0})$, etc.,
$m_{\tilde{t}_{L}} \approx m_{\tilde{t}_{R}}$,
and neglected all terms aside from $h_{t}$ and $g_{s}^{2}$
($ = 4\pi\alpha_{s}$) ones. ($M_{\tilde{g}}$ is the gluino mass.)
This is of course a rough estimate,
as, in practice all the parameters
scale with $Q$.
However, it
illustrates  qualitatively the presence
of residual two-loop  logarithms
which are enhanced
by the large couplings and masses
when evaluating $\Delta_{h^{0}}$.
Note that typically  the QCD term wins and the correction is
positive near $Q = M_{Z}$.

Though formally of two-loop order, the residual $Q$ dependence of the
EPM $m_{h^{0}}$ is (in our case) of order $10 - 20 \%$.
The ambiguity in $m_{h^{0}}$
is reduced for $m_{h^{0}}^{T} \gg 0$.
Also, for values of $\tan\beta$ outside the allowed region
$\Delta_{\theta_{t}}$
is diminished
(e.g., in the notation of
Ref. \cite{ez} only $\Delta_{22}$ is important),
$h_{t} < 1$ (which also diminishes the wave-function renormalization),
and thus, there is a smaller
ambiguity not only in $m_{h^{0}}$ but also in $\Delta_{h^{0}}$.
Nevertheless, one should be aware of such ambiguities, which
are generically  present in a calculation of that sort.

\subsection{The RGM: Diagonalize and run}

One could alternatively correct for thresholds by explicitly decoupling a
particle below its threshold
and redefining the effective field theory:
the renormalization group method (RGM).
Following Ref. \cite{haberdiaz,haberham},
we diagonalize
the Higgs mass matrices at scale $Q = m_{A^{0}}$,
integrate out the heavy Higgs doublet,
and define a one Higgs doublet model (1HDM)
effective theory with a quartic coupling
$\frac{\lambda}{2}h^{4}$.
Note that $m_{A^{0}} \gg m_{\tilde{t}}$, so only the heavy
Higgs doublet is integrated out at $Q = m_{A^{0}}$, and other
heavy particles will be decoupled at lower  thresholds.
The boundary condition for $\lambda(m_{A^{0}})$
is determined by the MSSM Higgs couplings
at that scale
and requires knowledge of $\tan\beta(m_{A^{0}})$
(i.e., wave-function renormalization).
Mixing effects are accounted for by a complicated
set of matching conditions
\cite{ham,haberham}.
$\lambda$ is then evolved down to $M_{Z}$, where
$m_{h^{0}} = \sqrt{\lambda}\nu
= 2\sqrt{\lambda}M_{W}/g_{2}$;
i.e., a ``diagonalize and run'' algorithm.

One can solve the RGE's iteratively to get
a semi-analytic approximation \cite{haberdiaz,haberham}.
This is not an exact procedure, and
at one-loop is accurate to
leading logarithms
(i.e., the accuracy of the EPM). Alternatively,
if the RGM calculation
is done by exact (or numerical)
integration, then the leading logarithms
are summed to all orders.
In agreement with
Ref. \cite{haberdiaz} we find that
$\lambda(M_{Z})$  (and thus $m_{h^{0}}$)
is diminished when integrating the RGE's numerically.
However, the effect of the numerical integration
(Fig.\ \ref{fig:f4})
is more important than anticipated in Ref. \cite{haberdiaz}
since $A^{0}$ is heavier than the mass-scale used there and
because of the large couplings.
The difference
between the numerical treatment and the approximate formula
is of order of two-loop (leading) logarithms.
The summation of the leading logarithms
modifies the scale dependence of the calculated mass.

A different scale dependence is via the details of the decoupling of
heavy states. To ease the calculation, we do not
diagonalize scalar-quark mass-squared matrices
at an intermediate scale. Instead, we put in the appropriate
matching conditions \cite{ham,haberham} by hand.
These conditions are
derived by expanding the EP assuming $|m_{t}\mu| \lesssim
m_{\tilde{t_{L}}}^{2}$, etc., which still holds in
our case [$m_{t}(m_{A^{0}}) < m_{t}(M_{Z})$].
Nevertheless, we
repeated the 1HDM RGM calculation but using
the EPM with $Q \approx m_{A^{0}}$ to determine
$m_{h^{0}}(m_{A^{0}})$, and thus [using $\nu(m_{A^{0}})$]
the boundary condition for
$\lambda(m_{A^{0}})$. The agreement with the calculation
using the boundary conditions of Ref. \cite{ham,haberham}
is good.

\subsection{Validity of the comparison between the EPM and RGM}

Whether ``running and diagonalizing''
or vice versa leads to a slightly different answer.
However,
it was already noted in Ref. \cite{haberdiaz} that the difference
between the two  algorithms is small and of higher order,
and can be ignored for the purpose of our discussion here.
However, the inclusion of the appropriate RGM
matching conditions is essential
for the comparison between the two methods.

More importantly, the RGM approximation of 1HDM breaks down if
$|\mu|$ is not sufficiently large.
(In such cases
the more conventional 2HDM RGM is appropriate.)
Too small scalar-quark (diagonal) masses break down the expansion of
Ref. \cite{ham,haberham} and too
large scalar threshold-corrections
can break down the EPM approximation.
Indeed, one has to ensure that the scalar quarks are not
heavier than about 1 TeV in order not to invalidate
the loop expansion which underlies the
EP, as these large logarithms would be  multiplied
by large couplings\footnote{This requirement
is consistent with our choice of ranges for the soft parameters.}
(see, for example, Ref. \cite{warsaw2}).
We checked that the points in the parameter space
where the RGM and EPM are in sharp disagreement
(e.g., in Figs.\ \ref{fig:f4} and \ref{fig:f5}) correspond to
either one of these cases.

\subsection{Two-loop calculations}

In this work we do not carry out a two-loop
calculation of $\Delta_{h^{0}}$. Such a calculation is an
elaborate task
and attention should be paid to various issues, e.g.,
wave-function renormalization and the coupled running
of the masses and couplings; pole masses $vs.$ running
masses; the multi-scale structure of the low-energy
theory; the correct choice of matching conditions (logarithmic
and non-logarithmic) given a choice of decoupling scales; one-loop
finite pieces (not included in the EP) that may become
important in certain regions of the parameter space; in addition to
correct counting of powers of the couplings. In fact,
an hybrid algorithm
between the EPM and RGM is required.

We received recently two interesting papers
where (partial) two-loop analyses were presented
\cite{twoloop,ham2}.
In this section we compare those results with
the residual two-loop terms we identified above.
We will take a somewhat critical point of view
emphasizing that though the above studies
(as well as the discussion here)
pave the way to understanding $\Delta_{h^{0}}$
beyond one-loop order, they are not complete.
We conclude that the one-loop leading-logarithm
result is most probably an upper bound on the two-loop result,
but we doubt whether any stronger conclusions can be drawn at
the present.

Above, we presented the
leading-logarithm result. In the EPM this is given by
$\Delta V^{one-loop}$ and in the RGM one
solves the RGE's to that order.
We found no significant
difference between the two methods. Next, we proceeded to
calculate beyond the leading order and integrated the
RGM RGE's numerically, summing the leading logarithms
to all orders in perturbation theory, and, in particular,
to two-loop order. We then found a $\sim -10-20\%$ correction
to $\Delta_{h^{0}}$. On the other hand, we identified
in the EPM expressions  next-to-leading two-loop
logarithmic terms. Those terms come about from the scale
dependence of the running parameters, and introduce
a similar (but positive) correction to
$\Delta_{h^{0}}$.
Also, we noted the important ($\sim 10\%$) effect
of $m_{t}^{pole}/m_{t}^{running} \approx [1 + (4\alpha_{s}/3\pi)]$
[e.g., $(m_{t}^{running})^{4} \approx 0.8(m_{t}^{pole})^{4}$].
It is then suggestive  that the two-loop
leading logarithms  and $m_{t}$ corrections
would diminish the one-loop
(leading-logarithm) result, while the next-to-leading
logarithms would partially counter-balance that effect.
Thus, we reach the conclusion that the one-loop leading-logarithm
prediction for $m_{h^{0}}$ (which contain some next-to-leading
logarithms) is an upper bound on the two-loop prediction.
Any stronger conclusion would require a detailed consideration
of the issues listed above. Furthermore, the significance
of a detailed two-loop study in the context of SGUT's
is not clear unless complemented by similar studies
of threshold effects near the high-scale boundary.

Let us proceed with the issue of the orders in
perturbation theory.
It was already pointed out by Chankowski \cite{warsaw2}
that the inclusion of $\Delta V^{one-loop}$
in (\ref{pot}) with only one-loop RGE's for the
(RG-improved) potential parameters is
an inconsistent procedure when counting
powers of couplings.
These issues have recently received great attention
in the context of theories with a single scalar field
(i.e., $\phi^{4}$ or the SM) \cite{phi4,bando3}.
It was explicitly shown that a RG-improved
L-loop effective potential is consistent when the
parameters are calculated using (L+1)-loop RGE's \cite{bando3}.

The  observations of Ref. \cite{phi4,bando3}
were recently applied to the MSSM by
Kodaira et al. \cite{twoloop} who extended
a previous work of Espinosa and Quiros \cite{madrid2}.
Indeed, they find that the two-loop RGM with one-loop EP
slightly increase the one-loop RGM result.
However, they impose the naive assumption of only one relevant
scale ($M_{SUSY}$) below which the SM RGE's are in effect
(and the EP contain only a $m_{t}$ correction).
Thus, the MSSM parameters are not treated on the same
footing as the SM ones, and
one expects some modification of their numerical results.
However, we note that
the correction is positive:
recall that the difference between
the two RGM approximations is the next-to-leading logarithms,
which we found to be positive by examining
the scale dependence of the result.

Hempfling and Hoang \cite{ham2} explicitly calculate the
two-loop RGM correction (using the methods of Ref.
\cite{haberham}), finding
\begin{equation}
\delta\Delta_{h^{0}}^{2} \approx  \frac{3}{(16\pi^{2})^{2}}
8\sqrt{2}G_{\mu}m_{t}^{4}\ln{\frac{M_{SUSY}^{2}}{m_{t}^{2}}}
\left[\frac{16}{3}g_{s}^{2} - 5h_{t}^{2}\right],
\label{ham}
\end{equation}
which roughly agrees with our Eq. (\ref{q}).

Those  authors go further to derive a diagramatic two-loop
result (in a simplified model)
which diminishes $\Delta_{h^{0}}$
below its  two-loop RGM prediction.
They then show that, e.g., if the difference between
the pole and
running $t$-quark mass is accounted for in the RGM calculation
by appropriate matching conditions, the two methods agree.
This again stresses the need for adequate choice of matching conditions.
However, it is not clear that there are no other effects
of that order of magnitude that can slightly increase $\Delta_{h^{0}}$.
[Also, we showed that scalar mixings which are neglected
in the above two-loop calculations and in our Eq. (\ref{q}),
are not apriori
negligible.]

In short, all analyses support the conclusion that two-loop
logarithms will diminish the one-loop
leading logarithm result (in particular, if one properly
distinguishes the $t$-quark pole mass from the running mass).
However,
it is not clear that the prediction of a
complete two-loop analysis lies
as low as suggested in Ref. \cite{ham2}
(or by our Fig.\ \ref{fig:f9}).
(If indeed this is the case our upper
bounds will be significantly strengthened.)

\section{Extended Higgs sectors}
\label{sec:app3}

If there is an extended heavy Higgs sector (as is assumed by some
authors when trying to explain the light fermion spectrum)
with large representations coupling to the third family
(i.e., there is no flavor symmetry that
forbids such couplings) then (\ref{yuk}) does not hold and
our constraints are evaded.
Alternatively, some models extend the light Higgs sector by adding a SM
singlet superfield, $S$, to the spectrum. This
is the only addition to the Higgs sector which is consistent
(i.e., without fine-tuned cancellations)
with
coupling constant unification.

\begin{figure}
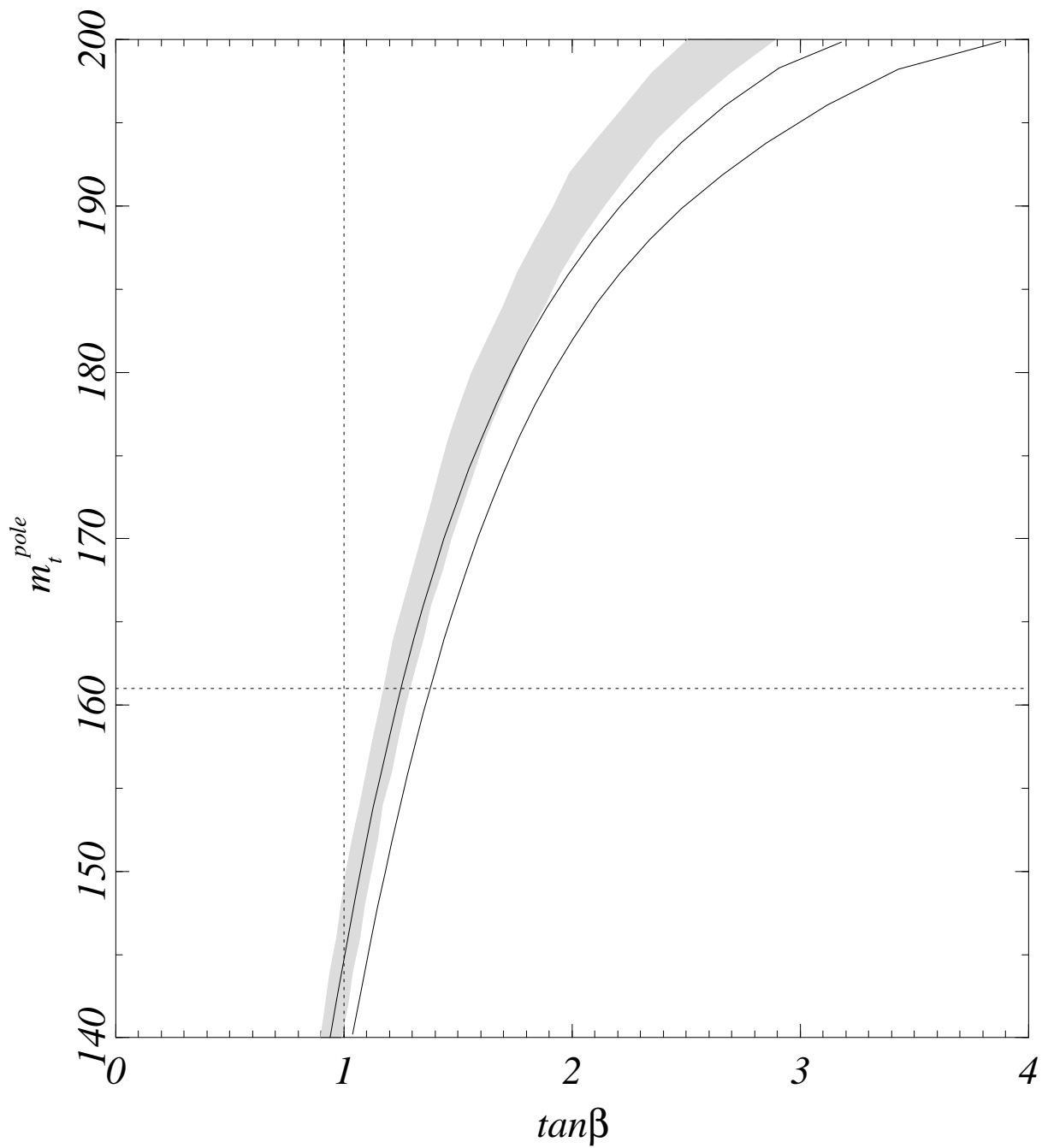

\caption{
The low $\tan\beta$ branch in the presence
of an additional Higgs field with
$\lambda_{s} = 0.5$ and $\kappa_{s} = 0$.
The allowed area when there is no singlet
field ($\lambda_{s} = 0$) is shown for comparison
(shaded region). the $t$-quark pole mass (given in GeV)
range suggested by precision data
and $\tan\beta = 1$ are indicated (dotted lines).
}
\label{fig:f10}
\end{figure}

The new singlet(s) would mix with
the other neutral Higgs bosons and would thus contribute
to $m_{h^{0}}^{T}$; e.g., \cite{slimit},
\begin{equation}
{m_{h^{0}}^{T}}^{2} \leq M_{Z}^{2}\cos^{2}2\beta
+ \lambda_{s}^{2}\frac{\nu^{2}}{2}\sin^{2}2\beta,
\label{singlet}
\end{equation}
where
$\lambda_{s}$
($\kappa_{s}$) is a new Yukawa coupling
in the superpotential: $\lambda_{s}H_{1}H_{2}S$
($\kappa_{s}S^{3}$). (We do not distinguish here
our notation of superfields from that of their scalar components.)
The MSSM limit with $m_{h^{0}}^{T} = 0$ is recovered
if $\lambda_{s} \rightarrow 0$,
which is indeed the case if
$h_{t} \approx h_{t}^{fixed}$ \cite{soham}.
However, the new Yukawa couplings (i.e., $\lambda_{s} \neq 0$)
enable one to realize (\ref{yuk})
for somewhat smaller values of $h_{t} < h_{t}^{fixed}$
(and $\tan\beta > 1$).
The shift in the allowed region is shown
in Fig.\ \ref{fig:f10} for $\lambda_{s} = 0.5$ and $\kappa_{s} = 0$.
Thus, $m_{h^{0}}^{T}$ can be large and $m_{h^{0}}$
can be heavy $\sim 150$ GeV \cite{madrid,mich1,soham}
when loop corrections are added. (See also Ref. \cite{sohamnew}.)

\begin{table}
\caption{The allowed regions in the $\tan\beta - m_{t}^{pole}$ plane.
Relations $(1)$ and $(2)$ are
also consistent with the parameter space described
by case $(2)$
except that $h_{t} \geq h_{b}$.
$m_{\tilde{q}}$ stands for a typical scalar-quark mass.
Finite corrections $\propto \tan\beta M_{\frac{1}{2}}\mu/m_{0}^{2}$
could modify the allowed $m_{t}^{pole}$ range in case $(2)$.
}
\label{table:t1}
\begin{tabular}{ c c c}
 &
case $(1)$ &
case $(2)$ \\
\hline
\hline
 GUT relations &
(\ref{cc}) and (\ref{yuk}) &
(\ref{cc}) and (\ref{yuk2}) \\

The allowed $m_{t}^{pole}$ range (GeV) &
$140  -  \sim 190 $ &
$\sim 180 \pm 15$ \\

The allowed $\tan\beta$ range &
$1 - 2$ &
$50 \pm 10$ \\

Yukawa couplings &
$h_{t} \approx 1 \gg h_{b}$ &
$h_{t} \approx  h_{b} \approx 1 $\\

The Higgs potential (approximate) symmetry &
$SU(2)_{L} \times SU(2)_{R}$ &
$O_{4} \times O_{4}$ \\

The vacuum (approximate) symmetry &
$SU(2)_{L + R} $ &
$O_{3} \times O_{3}$ \\

$SU(2) \times U(1)$ breaking ``messenger''&
$m_{3}^{2}$ &
$m_{H_{2}}^{2}$ \\

The large mass parameter&
Higgsino mass parameter &
scalar-quark masses \\

$m_{h^{0}}^{T}$ &
$m_{h^{0}}^{T} \approx 0$ &
$m_{h^{0}}^{T} \approx M_{Z}$ \\

$\Delta_{h^{0}}$ enhancement&
left-right mixings&
large $m_{t}$ and the heavy scalars \\

$m_{H^{0}}$ &
$m_{H^{0}} \approx m_{A^{0}}$ &
$m_{H^{0}} \approx m_{A^{0}}$ \\

$A^{0}$, $H^{+}$ &
\begin{tabular}{c}
heavy\\
$m_{A^{0}} \approx \sqrt{2}|\mu| \approx 2$ TeV\\
\end{tabular}&
\begin{tabular}{c}
massive pseudo-Goldstone bosons \\
$m_{A^{0}} \ll m_{\tilde{q}}$\\
\end{tabular}\\

\end{tabular}
\end{table}

\end{document}